\documentstyle[preprint,epsf]{jpsj}

\def\runtitle{
Roles of Electron Correlations 
in the Spin-Triplet Superconductivity of Sr$_2$RuO$_4$}
\def\runauthor{Takuji {\sc Nomura}}

\title{
Roles of Electron Correlations\\ 
in the Spin-Triplet Superconductivity of Sr$_2$RuO$_4$ 
}

\author{
Takuji {\sc Nomura} and Kosaku Yamada
}

\inst{
Department of Physics, Kyoto University, Kyoto, 606-8502, Japan
}

\recdate
{
}

\abst{
We discuss a microscopic mechanism of the spin-triplet superconductivity 
in the quasi-two-dimensional ruthenium oxide Sr$_2$RuO$_4$ 
on the basis of two-dimensional three-band Hubbard model. 
We solve the linearized \'Eliashberg equation 
by taking into account the full momentum-frequency 
dependence of the order parameter 
for the spin-triplet and the spin-singlet states, 
and estimate the transition temperature as a function 
of the Coulomb integrals. 
The effective pairing interaction is expanded perturbatively 
with respect to the Coulomb interaction at the Ru sites up to 
the third order. As a result, we show that the spin-triplet 
$p$-wave state is more stable than the spin-singlet $d$-wave state 
for moderately strong Coulomb interaction. 
Our results suggest that one of the three bands, $\gamma$, 
plays a dominant role in the superconducting transition, 
and the pairing on the other two bands($\alpha$ and $\beta$) 
is induced passively through the inter-orbit couplings. 
The most significant momentum dependence for the $p$-wave pairing 
originates from the vertex correction terms, 
while the incommensurate antiferromagnetic spin fluctuations, 
which are observed in inelastic neutron scattering experiments, 
are expected to disturb the $p$-wave pairing 
by enhancing the $d$-wave pairing. 
Therefore we can regard the spin-triplet superconductivity 
in Sr$_2$RuO$_4$ as one of the natural results of the electron correlations, 
but cannot consider as a result of some strong magnetic fluctuations. 
We will also mention the normal Fermi liquid properties of Sr$_2$RuO$_4$. 
}

\kword{
electron correlations, three-band Hubbard model, 
third order perturbation theory, vertex corrections, 
Sr$_2$RuO$_4$, spin-triplet superconductivity, 
superconducting transition temperature. 
}

\begin{document}
\sloppy
\maketitle

\section{Introduction}
\label{Sc:Intro}

Clarifying the mechanism of the spin-triplet 
superconductivity in Sr$_2$RuO$_4$ has been 
one of the most challenging issues 
in the physics of strongly correlated electrons
~\cite{Rf:Maen.1,Rf:Maen.2}. 
This unconventional pairing state has been accepted 
widely as a result of many intensive experimental 
and theoretical studies. 

Some of the most remarkable results supporting 
the odd-parity parallel-spin pairing were obtained 
by the direct measurements of the spin susceptibility 
through the superconducting transition. 
The NMR Knight shift~\cite{Rf:Ishi.1} 
and the inelastic polarized neutron~\cite{Rf:Duff.1} 
scattering measurements revealed that 
the spin susceptibility of the electron system 
does not show any change at the transition. 
These facts indicate that the electron system does not 
lose all the activity in the spin space 
even in the superconducting phase, 
and exclude the possibility of the singlet Cooper pairing. 
In addition, there are many other observations 
suggesting indirectly but very strongly the realization 
of the unconventional pairing symmetry. 
The superconductivity is crucially destroyed 
by the slight concentration of nonmagnetic impurities~\cite{Rf:Mack.1}, 
in contrast to the conventional $s$-wave superconductivity. 
The NMR relaxation rate exhibits no coherence 
peak just below $T_{\rm c}$~\cite{Rf:Ishi.2}. 
The muon spin relaxation rate is increased below $T_{\rm c}$, 
which indicates that an internal magnetic field is turned on 
spontaneously and the time-reversal symmetry breaks down 
in the superconducting state~\cite{Rf:Luke.1}. 
More recently there are several experiments reporting 
that the second superconducting phase transition occurs 
in the vicinity of $H_{{\rm c}_2}$ curve~\cite{Rf:Mao.1,Rf:Nish.1}. 

The normal state properties of Sr$_2$RuO$_4$ are 
quite different from those of high-$T_{\rm c}$ cuprates, 
in spite of the similarity in the crystal structures. 
Sr$_2$RuO$_4$ possesses a metallic conductivity 
without any carrier doping, and exhibits typical 
quasi-two-dimensional Fermi liquid behaviors, 
for example, the Pauli paramagnetism, the $T^2$ behavior 
in the resistivity at low temperatures~\cite{Rf:Maen.1, Rf:Maen.3}. 
The electronic structure is characterized 
by the three cylindrical Fermi surfaces, 
$\alpha$, $\beta$ and $\gamma$, 
according to the de Haas-van Alphen measurements~\cite{Rf:Mack.2}. 
This is in good agreement with the results 
of the band structure calculations~\cite{Rf:Oguc.1,Rf:Sing.1}. 

From the theoretical point of view, Rice and Sigrist 
predicted the parallel-spin pairing shortly after the 
discovery of the superconductivity~\cite{Rf:Rice.1}. 
Their discussions are based on the facts that 
Sr$_2$RuO$_4$ gives a set of Fermi liquid parameters 
comparable to those of the superfluid $^3$He and 
that the three-dimensional version SrRuO$_3$ is a ferromagnet. 
According to that situation, it seems quite natural to expect 
that some ferromagnetic spin fluctuations induce 
the spin-triplet superconductivity~\cite{Rf:Mazi.1,Rf:Mont.1}. 
However the inelastic neutron scattering measurements did not 
detect any enhancement of ferromagnetic components 
but observed a sizeable incommensurate antiferromagnetic fluctuations 
in the spin susceptibility $\chi(q)$~\cite{Rf:Sidi.1}. 
Therefore it is difficult to expect that 
some strong ferromagnetic fluctuations play the leading role 
for the superconductivity of Sr$_2$RuO$_4$. 

Here we would like to comment on some other 
theoretical considerations on the mechanism of the superconductivity. 
Some theorists consider that the observed incommensurate 
antiferromagnetic fluctuations mediate 
the superconductivity~\cite{Rf:Kuwa.1,Rf:Kuro.1}. 
According to the discussions, the incommensurate fluctuations 
should possess a large anisotropy, 
$\chi_{zz}(Q_{\rm inc}) > 2 \chi_{-+}(Q_{\rm inc})$. 
To our knowledge, however, such a high anisotropy has never been 
observed in Sr$_2$RuO$_4$. 
Particularly, the recent neutron scattering measurement 
by Servant {\it et al.} suggests that the spin susceptibility 
$\chi(Q_{\rm inc})$ is very isotropic~\cite{Rf:Serv.1}. 
Recently magnetic properties of Ti-doped Sr$_2$Ru$_{1-x}$Ti$_x$O$_4$
have been investigated~\cite{Rf:Mina.1}. 
This compound exhibits the long range magnetic ordering for $x \geq 0.03$, 
which evolves from the incommensurate antiferromagnetic fluctuations.
In the ordered phase, the spin moment points 
along the $c$-axis direction~\cite{Rf:Brad.1}. 
This anisotropy is qualitatively identical to that required 
in the anisotropic spin fluctuation scenario. 
What we should note here is, however, the following point. 
The effect of Ti-doping on the superconducting transition 
temperature is quite similar to that of other nonmagnetic 
impurity doping, in spite that the incommensurate fluctuations 
are much enhanced by Ti-doping~\cite{Rf:Kiku.1}. 
If the incommensurate fluctuations mediated the superconductivity, 
the transition temperature should behave for Ti-dopings 
in a different way from the case of other nonmagnetic impurity doping. 
Consequently, we consider that the incommensurate 
antiferromagntetic fluctuations are not essential for the 
superconductivity of Sr$_2$RuO$_4$. 

A scenario based on the incommensurate orbital fluctuations was 
discussed~\cite{Rf:Taki.1}. 
In order to explain the spin-triplet 
superconductivity, however, they require rather strong 
inter-orbit repulsion between Ru4d$\varepsilon$ electrons. 
The effect of the inter-orbit repulsion 
usually becomes weak remarkably by the screening. 
Therefore we consider that the incommensurate orbital fluctuations 
are not large enough to stabilize the spin-triplet pairing 
in Sr$_2$RuO$_4$. 

There are some other scenarios based 
on the Hund's coupling~\cite{Rf:Bask.1,Rf:Ng.1,Rf:Spal.1}. 
The Hund's coupling is usually attractive only 
for two parallel-spin particles 
which are in the different orbits on the same localized atom. 
Therefore it seems that the Hund's coupling does not necessarily play 
an important part for stabilizing the parallel spins 
over the coherence length. 
In addition, the hybridization among the atomic orbitals Ru4d$\varepsilon$ 
is relatively small in Sr$_2$RuO$_4$, because of both 
the two-dimensionality and the symmetry of the atomic wave functions. 
We consider that the Hund's coupling is actually hard to be the main origin 
of the spin-triplet superconductivity in Sr$_2$RuO$_4$. 

Recently the present authors discussed on the basis of the one-band 
Hubbard model for the single branch $\gamma$~\cite{Rf:Nomu.1}. 
According to the previous works~\cite{Rf:Nomu.1, Rf:Nomu.2}, 
the spin-triplet superconductivity 
of Sr$_2$RuO$_4$ is a natural result of the electron correlations. 
Even if the spin susceptibility has no prominent peak 
around ${\mib q}=0$, the effective pairing interaction 
has a momentum dependence favorable for the $p$-wave pairing state.
The aim of the present article is to extend the previous 
single-band discussion to three-band one and give 
a comprehensive understanding on the mechanism of 
the $p$-wave pairing. 
As a result, we show the following points: the spin-triplet 
$p$-wave state is more stable than the spin-singlet $d$-wave state 
for moderately strong Coulomb interaction. 
One of the three bands, $\gamma$, plays the dominant role 
in the superconducting transition, 
and the pairing on the other two bands($\alpha$ and $\beta$) 
is induced passively through the inter-orbit couplings. 
The most significant momentum dependence for the $p$-wave pairing 
originates from the vertex correction terms, 
and the incommensurate antiferromagnetic spin fluctuations 
are expected to disturb the $p$-wave pairing 
by enhancing the $d$-wave pairing. 
Therefore we can regard the spin-triplet superconductivity 
in Sr$_2$RuO$_4$ as one of the natural results of electron correlations, 
and cannot consider as a result of some strong magnetic fluctuations. 

This paper is organized as follows. 
In \S~\ref{Sc:Formulation}, 
we give the formulation including introduction of 
the Hubbard Hamiltonian and perturbation expansion. 
In \S~\ref{Sc:Results}, 
we show the results of the numerical calculations 
of $T_{\rm c}$ as a function of the Coulomb energy, 
anomalous self-energy and effective interaction. 
There we mention also normal Fermi liquid properties, 
Fermi surface, normal self-energy and density of states. 
Finally, in \S~\ref{Sc:DiscConc}, 
we comment on the results, give a proposal for experiments, 
and conclude the paper. 

\section{Formulation}
\label{Sc:Formulation}

\subsection{Three-band Hubbard model}
\label{Sc:Model}

Sr$_2$RuO$_4$ is one of typical strongly correlated electron systems 
where Ru4d electrons play the most significant role 
for the electronic properties. 
According to the band calculation results, 
Ru4d$\varepsilon$ orbitals take the main part of 
the density of states near the Fermi level, 
although the Ru4d$\varepsilon$ and O2p orbitals 
are hybridizing~\cite{Rf:Oguc.1,Rf:Sing.1}. 
Therefore the electronic structure is expected to be 
reproduced well by considering three Wannier states, $xy$, $yz$ and $xz$, 
which have the characters of Ru4d$_{xy, yz, xz}$. 

We give the Hamiltonian in the form, 
\begin{equation}
H=H_0+H', 
\end{equation}
where $H_0$ and $H'$ are the noninteracting part 
and interaction part, respectively.
$H_0$ is given by 
\begin{equation}
H_0 = \sum_{{\mib k},\ell,\sigma}{\xi}_{\ell}({\mib k})
c_{{\mib k}\ell\sigma}^{\dagger}c_{{\mib k}\ell\sigma}
+ \sum_{{\mib k},{\sigma}}\lambda({\mib k})
(c_{{\mib k}\,yz\,\sigma}^{\dagger}c_{{\mib k}\,xz\,\sigma}
+c_{{\mib k}\,xz\,\sigma}^{\dagger}c_{{\mib k}\,yz\,\sigma}), 
\end{equation}
where $c^{\dagger}$($c$) is electron creation(annihilation) operator, 
and $\mib{k}$, ${\sigma}^{(}{}'^{)}$ and $\ell$ denote momentum, 
spin and the Wannier states $(xy,yz,xz)$, respectively.
We take the following band dispersions,
\begin{eqnarray}
{\xi}_{xy}({\mib k}) &= &2t_1({\cos}k_x+{\cos}k_y)
+ 4t_2{\cos}k_x{\cos}k_y-\mu_{xy},\\ 
{\xi}_{yz}({\mib k}) &= &2t_3{\cos}k_y+2t_4{\cos}k_x-\mu_{yz},\\ 
{\xi}_{xz}({\mib k}) &= &2t_3{\cos}k_x+2t_4{\cos}k_y-\mu_{xz},\\ 
\lambda({\mib k}) &= &4{\lambda}_0{\sin}k_x{\sin}k_y.
\end{eqnarray}
$t_1$, $t_2$, $t_3$, $t_4$ and $\lambda_0$ are transfer integrals
between nearest or next nearest neighbor Ru sites (Fig.~\ref{Fg:model}), 
where we have assumed tetragonal symmetry, and 
$\mu_{\ell}$ is the chemical potential. 
For the present calculation, 
we take $t_1=-1.00$, $t_2=-0.400$, $t_3=-1.25$, $t_4=-0.125$ and 
$\lambda_0=-0.200$, and determine the chemical potentials $\mu_{\ell}$ 
to satisfy $n_{xy}=0.700$, $n_{yz}=0.700$ and $n_{xz}=0.700$, 
where $n_{\ell}$ is the electron filling per one spin state 
in the orbital $\ell$. 
The total population of the conduction electrons at a Ru site 
equals $2(n_{xy}+n_{yz}+n_{xz})=4.200$, 
which is a little larger than  the electron number 
obtained by the experiments~\cite{Rf:Mack.2}, 4.032. 
We discuss this deviation in \S~\ref{Sc:DiscConc}. 
We take the following on-site Coulomb interaction for 
the interacting part, 
\begin{eqnarray}
H' &= &\frac{1}{2}U\sum_{i}\sum_{\ell}\sum_{\sigma\neq\sigma'}
c_{i\ell\sigma}^{\dagger}c_{i\ell\sigma'}^{\dagger}
c_{i\ell\sigma'}c_{i\ell\sigma}
+ \frac{1}{2}U'\sum_{i}\sum_{\ell\neq\ell'}\sum_{\sigma,\sigma'}
c_{i\ell\sigma}^{\dagger}c_{i\ell'\sigma'}^{\dagger}
c_{i\ell'\sigma'}c_{i\ell\sigma}\nonumber\\
&+ &\frac{1}{2}J\sum_{i}\sum_{\ell\neq\ell'}\sum_{\sigma,\sigma'}
c_{i\ell\sigma}^{\dagger}c_{i\ell'\sigma'}^{\dagger}
c_{i\ell\sigma'}c_{i\ell'\sigma}
+ \frac{1}{2}J'\sum_{i}\sum_{\ell\neq\ell'}\sum_{\sigma\neq\sigma'}
c_{i\ell\sigma}^{\dagger}c_{i\ell\sigma'}^{\dagger}
c_{i\ell'\sigma'}c_{i\ell'\sigma}. 
\end{eqnarray}
$J$ is the Hund's coupling. 
For the convenience of calculation, 
we express the Coulomb integrals by a matrix 
$I_{(\ell_1\sigma_1)(\ell_2\sigma_2),(\ell_3\sigma_3)(\ell_4\sigma_4)}$, 
where the matrix $I$ has the following elements, 
\begin{equation}
\left.
\begin{array}{lll}
I_{(\ell\sigma)(\ell\bar{\sigma}),(\ell\bar{\sigma})(\ell\sigma)} &= &U\\
I_{(\ell\sigma)(\ell'\sigma),(\ell'\sigma)(\ell\sigma)} &= &U'-J\\
I_{(\ell\sigma)(\ell'\bar{\sigma}),(\ell'\bar{\sigma})(\ell\sigma)} &= &U'\\
I_{(\ell\sigma)(\ell'\bar{\sigma}),(\ell\bar{\sigma})(\ell'\sigma)} &= &J\\
I_{(\ell\sigma)(\ell\bar{\sigma}),(\ell'\bar{\sigma})(\ell'\sigma)} &= &J'
\end{array}\right\},
\end{equation}
where $\ell\neq\ell'$, and $\sigma\neq\bar{\sigma}$, 
and the other elements are zero. 
Using the matrix $I$, $H'$ is expressed in the simple form, 
\begin{equation}
H'=\frac{1}{2}\sum_{i}\sum_{\zeta_1\zeta_2\zeta_3\zeta_4}
I_{\zeta_1\zeta_2,\zeta_3\zeta_4}
c_{i\zeta_1}^{\dagger}c_{i\zeta_2}^{\dagger}
c_{i\zeta_3}c_{i\zeta_4}, 
\end{equation}
where we have adopted the short notations, $\zeta=(\ell\sigma)$ 
and $\sum_{\zeta}=\sum_{\ell}\sum_{\sigma}$. 
We define the antisymmetric bare vertex by 
\begin{equation}
\Gamma^{(0)}_{\zeta_1\zeta_2,\zeta_3\zeta_4}
=I_{\zeta_1\zeta_2,\zeta_3\zeta_4}-I_{\zeta_1\zeta_2,\zeta_4\zeta_3}.
\end{equation}
In the perturbation expansion, this vertex is represented diagrammatically 
by the empty square as depicted in Fig.~\ref{Fg:bvert}. 
If we have in mind the rotation invariance of the d$\varepsilon$ 
atomic orbital wave functions in orbital space, 
the Coulomb integrals, $U$, $U'$, $J$ and $J'$, are 
not independent parameters, and the relations $U=U'+2J$ and $J=J'$ 
are derived~\cite{Rf:Dagotto.1}. 
We cannot, however, easily tell that the relation should 
be valid even for the Wannier wave functions in metallic solids, 
because the Coulomb interaction may be changed effectively, 
in general, for example, due to the screening effect.     
In the present calculation, therefore we treat the Coulomb integrals 
as independent parameters. 
In any case, as we can see in \S \ref{Sc:Egnvls}, 
we obtain the $p$-wave superconductivity 
for a broad parameter region of the inter-orbit couplings, 
and may conclude that this assumption of the independence 
is not significant in the present discussions. 

In the previous work, 
we investigated the magnetic properties of layered ruthenates 
on the basis of the same model~\cite{Rf:Nomu.3}. 

\subsection{Perturbation expansions}
\label{Sc:Expansion}

We expand the normal self-energy and the effective pairing interaction 
perturbatively up to the third order with respect 
to the Coulomb interaction $H'$. 
At first, we neglect the hybridizing term including $\lambda(\mib{k})$.
In Sr$_2$RuO$_4$, $\lambda(\mib{k})$ is the second nearest hopping term 
and expected to be small compared with the band width. 
By this simplification, the normal self-energy 
and the effective interaction are expanded perturbatively 
by the bare Green's functions, 
\begin{equation}
G^{(0)}_{\ell}(k)=\frac{1}{{\rm i}\omega_n - \xi_{\ell}({\mib k})}, 
\end{equation}
where $\ell = \{xy, yz, xz\}$ and $k=(\mib{k}, {\rm i}\omega_n)$. 

In the present scheme, the normal self-energy has 
only the diagonal elements $\Sigma_{\ell}(k)$. 
The perturbation terms to be summed up are 
displayed in Fig.~\ref{Fg:nself}. 
Since the terms represented by Fig.~\ref{Fg:nself}(N1) 
only shift the chemical potentials $\mu_{\ell}$, we regard them 
as already included in $\mu_{\ell}$'s. 
The general forms of perturbation terms remained for summations 
are given as follows.
\begin{equation}
\Sigma_{\zeta_1\zeta_2}^{\rm N2}(k) = 
- \frac{1}{2}\Bigl(\frac{T}{N}\Bigr)^2
\sum_{k_1,k_2}\sum_{\zeta_3\zeta_4}
\sum_{\xi_1\xi_2\xi_3\xi_4}
G_{\zeta_4\zeta_3}^{(0)}(k_1)G_{\xi_4\xi_2}^{(0)}(k_2)
\times G_{\xi_3\xi_1}^{(0)}(k-k_1+k_2)
\Gamma_{\zeta_1\xi_2\xi_3\zeta_4}^{(0)}
\Gamma_{\xi_1\zeta_3\zeta_2\xi_4}^{(0)}, 
\end{equation}
\begin{eqnarray}
\Sigma_{\zeta_1\zeta_2}^{\rm N3a}(k)
&= &- \Bigl(\frac{T}{N}\Bigr)^3
\sum_{k_1,k_2,k_3}\sum_{\zeta_3\zeta_4}
\sum_{\xi_1\xi_2\xi_3\xi_4}
\sum_{\gamma_1\gamma_2\gamma_3\gamma_4}
G_{\zeta_4\zeta_3}^{(0)}(k_1)
G_{\xi_4\xi_2}^{(0)}(k_2)G_{\xi_3\xi_1}^{(0)}(k-k_1+k_2)\nonumber\\
&&\times G_{\gamma_4\gamma_2}^{(0)}(k_3)G_{\gamma_3\gamma_1}^{(0)}(k-k_1+k_3)
\Gamma_{\zeta_1\xi_2\xi_3\zeta_4}^{(0)}
\Gamma_{\xi_1\gamma_2\gamma_3\xi_4}^{(0)}
\Gamma_{\gamma_1\zeta_3\zeta_2\gamma_4}^{(0)},
\end{eqnarray}
\begin{eqnarray}
\Sigma_{\zeta_1\zeta_2}^{\rm N3b}(k)
&= &\frac{1}{4}\Bigl(\frac{T}{N}\Bigr)^3
\sum_{k_1,k_2,k_3}\sum_{\zeta_3\zeta_4}
\sum_{\xi_1\xi_2\xi_3\xi_4}
\sum_{\gamma_1\gamma_2\gamma_3\gamma_4}
G_{\zeta_3\zeta_4}^{(0)}(k_1)
G_{\xi_3\xi_2}^{(0)}(k_2)G_{\xi_4\xi_1}^{(0)}(k+k_1-k_2)\nonumber\\
&&\times G_{\gamma_3\gamma_2}^{(0)}(k_3)G_{\gamma_4\gamma_1}^{(0)}(k+k_1-k_3)
\Gamma_{\zeta_1\zeta_4\xi_3\xi_4}^{(0)}
\Gamma_{\xi_1\xi_2\gamma_3\gamma_4}^{(0)}
\Gamma_{\gamma_1\gamma_2\zeta_3\zeta_2}^{(0)}.
\label{Eq:N3b}
\end{eqnarray}
The factors $\frac{1}{2}$ and $\frac{1}{4}$ multiplied
in the terms $\Sigma_{\zeta_1\zeta_2}^{\rm N2}(k)$ and 
$\Sigma_{\zeta_1\zeta_2}^{\rm N3b}(k)$
are necessary to avoid the redundancy in the summation.
By the simplification mentioned above, 
\begin{equation}
G^{(0)}_{\zeta\zeta'}(k)=G^{(0)}_{\ell}(k)\delta_{\zeta\zeta'}, 
\end{equation}
where $\zeta = (\ell\sigma)$, $\zeta' = (\ell'\sigma')$, 
and $\delta_{\zeta\zeta'}=\delta_{\ell\ell'}\delta_{\sigma\sigma'}$. 
The diagonal elements of the normal self-energy are reduced to 
\begin{eqnarray}
\Sigma_{\ell}(k) &= &\frac{T}{N}\sum_{k_1}\sum_{\ell_2}
G^{(0)}_{\ell_2}(k_1)\Bigl[
\frac{1}{2}\sum_{{m_1}{m_2}}X^{(0)}_{{m_1}{m_2}}(k-k_1)
\sum_{\nu_1\nu_2}\sum_{\sigma_2}
\Gamma^{(0)}_{(\ell\sigma)({m_2}\nu_2),({m_1}\nu_1)(\ell_2\sigma_2)}
\Gamma^{(0)}_{({m_1}\nu_1)(\ell_2\sigma_2),(\ell\sigma)({m_2}\nu_2)}\nonumber\\
&&-\sum_{{n_1}{n_2}}\sum_{{m_1}{m_2}}
X^{(0)}_{{n_1}{n_2}}(k-k_1)X^{(0)}_{{m_1}{m_2}}(k-k_1)
\sum_{\tau_1\tau_2}\sum_{\nu_1\nu_2}\sum_{\sigma_2}
\Gamma^{(0)}_{(\ell\sigma)({n_2}\tau_2),({n_1}\tau_1)(\ell_2\sigma_2)}\nonumber\\
&&\times
\Gamma^{(0)}_{({n_1}\tau_1)(m_2\nu_2),({m_1}\nu_1)({n_2}\tau_2)}
\Gamma^{(0)}_{({m_1}\nu_1)(\ell_2\sigma_2),(\ell\sigma)({m_2}\nu_2)}\nonumber\\
&&-\frac{1}{4}\sum_{{n_1}{n_2}}\sum_{{m_1}{m_2}}
\Phi^{(0)}_{{n_1}{n_2}}(k+k_1)\Phi^{(0)}_{{m_1}{m_2}}(k+k_1)
\sum_{\tau_1\tau_2}\sum_{\nu_1\nu_2}\sum_{\sigma_2}
\Gamma^{(0)}_{(\ell\sigma)({\ell_2}\sigma_2),({n_1}\tau_1)(n_2\tau_2)}\nonumber\\
&&\times
\Gamma^{(0)}_{({n_1}\tau_1)(n_2\tau_2),({m_1}\nu_1)({m_2}\nu_2)}
\Gamma^{(0)}_{({m_1}\nu_1)({m_2}\nu_2),(\ell\sigma)({\ell_2}\sigma_2)}
\Bigr],\label{Eq:N3}
\end{eqnarray}
where the functions, 
$X^{(0)}_{\ell\ell'}(q)$ and $\Phi^{(0)}_{\ell\ell'}(q)$, 
are calculated by 
\begin{eqnarray}
X^{(0)}_{\ell\ell'}(q) &=& -\frac{T}{N}\sum_{k}
G^{(0)}_{\ell}(q+k)G^{(0)}_{\ell'}(k),\label{Eq:Chi0}\\
\Phi^{(0)}_{\ell\ell'}(q) &=& -\frac{T}{N}\sum_{k}
G^{(0)}_{\ell}(q-k)G^{(0)}_{\ell'}(k).\label{Eq:Phi0}
\end{eqnarray}

The renormalized Green's functions are given by 
\begin{equation}
G_{\ell}(k)=\frac{1}{{\rm i}\omega_n - \Xi_{\ell}(k)}, 
\end{equation}
where $\Xi_{\ell}(k)
=\xi_{\ell}({\mib k})+\Sigma_{\ell}(k)-\delta \mu_{\ell}$.
$\delta \mu_{\ell}$ is determined to satisfy the relation
\begin{equation}
\delta n _{\ell}=\frac{T}{N}\sum_{k}(G_{\ell}(k)-G^{(0)}_{\ell}(k))=0.
\end{equation}

Here we take into account the effect 
of the hybridizing term $\lambda(\mib{k})$. 
We consider the following renormalized Green's functions: 
\begin{eqnarray}
G_{xy, xy}(k) &=& \frac{1}{{\rm i}\omega_n - \Xi_{xy}(k)},\\
G_{yz, yz}(k) &=& \frac{1}{{\rm i}\omega_n - \Xi_{yz}(k)
-\frac{\lambda^2(\mib{k})}{{\rm i}\omega_n - \Xi_{xz}(k)}},\\
G_{xz, xz}(k) &=& \frac{1}{{\rm i}\omega_n - \Xi_{xz}(k)
-\frac{\lambda^2(\mib{k})}{{\rm i}\omega_n - \Xi_{yz}(k)}},\\
G_{yz, xz}(k) &=& G_{xz, yz}(k) \\
&=&\frac{\lambda(\mib{k})}{({\rm i}\omega_n - \Xi_{yz}(k))
({\rm i}\omega_n - \Xi_{xz}(k))-\lambda^2(\mib{k})},\nonumber
\end{eqnarray}
and the other elements of $G_{\ell\ell'}(k)$ are zero.
These Green's functions are the matrix elements of
the operator $[i\omega_n - H_0 - (\Sigma-\delta\mu)]^{-1}$,
where $\Sigma-\delta\mu$ is an operator
with the diagonal matrix elements
$(\Sigma_{\ell}(k)-\delta\mu_{\ell})\delta_{\ell\ell'}$. 
Diagonalizing this renormalized Green's function, 
we obtain the following form 
\begin{equation}
G_{\nu}(k)=\frac{1}{{\rm i}\omega_n - \Xi_{\nu}(k)}. 
\end{equation}
Here $\nu=\{ \alpha, \beta, \gamma \}$, and 
\begin{eqnarray}
\Xi_{\alpha, \beta}(k) &= &\Xi_{+}(k) \mp [\Xi_{-}^2(k)
+\lambda^2(\mib{k})]^\frac{1}{2}\\
\Xi_{\pm}(k) &= &\frac{1}{2}(\Xi_{yz}(k) \pm \Xi_{xz}(k))\\
\Xi_{\gamma}(k) &=& \Xi_{xy}(k)
\end{eqnarray}

Then we expand the anomalous self-energy perturbatively 
in a similar manner. 
The perturbation terms to be summed up are displayed diagrammatically 
in Fig.~\ref{Fg:aself}. 
The general forms for the perturbation terms are given as follows. 
\begin{equation}
\Sigma_{\zeta_1\zeta_2}^{\rm A1\dag}(k) 
= -\frac{1}{2}\frac{T}{N}\sum_{k'}\sum_{\zeta_3\zeta_4}
F_{\zeta_4\zeta_3}^{\dag}(k')\Gamma_{\zeta_4\zeta_3,\zeta_2\zeta_1}^{(0)},
\end{equation}
\begin{equation}
\Sigma_{\zeta_1\zeta_2}^{\rm A2\dag}(k) = - \Bigl(\frac{T}{N}\Bigr)^2
\sum_{k',k_1}\sum_{\zeta_3\zeta_4}
\sum_{\gamma_1\gamma_2\gamma_3\gamma_4}
F_{\zeta_4\zeta_3}^{\dag}(k')G_{\gamma_4\gamma_3}^{(0)}(k_1)
G_{\gamma_2\gamma_1}^{(0)}(k+k_1-k')
\Gamma_{\zeta_4\gamma_3,\gamma_2\zeta_1}^{(0)}
\Gamma_{\gamma_1\zeta_3,\zeta_2\gamma_4}^{(0)},
\end{equation}
\begin{eqnarray}
\Sigma_{\zeta_1\zeta_2}^{\rm A3a\dag}(k) 
&= &- \Bigl(\frac{T}{N}\Bigr)^3
\sum_{k',k_1,k_2}\sum_{\zeta_3\zeta_4}
\sum_{\xi_1\xi_2\xi_3\xi_4}
\sum_{\gamma_1\gamma_2\gamma_3\gamma_4}
F_{\zeta_4\zeta_3}^{\dag}(k')
G_{\xi_4\xi_2}^{(0)}(k_1)G_{\xi_3\xi_1}^{(0)}(k-k'+k_1)\nonumber\\
&&\times G_{\gamma_4\gamma_2}^{(0)}(k_2)G_{\gamma_3\gamma_1}^{(0)}(k-k'+k_2)
\Gamma_{\zeta_4\xi_2,\xi_3\zeta_1}^{(0)}
\Gamma_{\xi_1\gamma_2,\gamma_3\xi_4}^{(0)}
\Gamma_{\gamma_1\zeta_3,\zeta_2\gamma_4}^{(0)},
\end{eqnarray}
\begin{eqnarray}
\Sigma_{\zeta_1\zeta_2}^{\rm A3b\dag}(k) 
&= & \Bigl(\frac{T}{N}\Bigr)^3
\sum_{k',k_1,k_2}\sum_{\zeta_3\zeta_4}
\sum_{\xi_1\xi_2\xi_3\xi_4}
\sum_{\gamma_1\gamma_2\gamma_3\gamma_4}
F_{\zeta_4\zeta_3}^{\dag}(k')
G_{\gamma_1\gamma_2}^{(0)}(-k+k'+k_1)G_{\gamma_3\gamma_4}^{(0)}(k_1)\nonumber\\
&&\times G_{\xi_3\xi_2}^{(0)}(-k+k_1+k_2)G_{\xi_4\xi_1}^{(0)}(k_2)
\Gamma_{\zeta_4\gamma_2,\gamma_3\zeta_1}^{(0)}
\Gamma_{\gamma_4\xi_1,\zeta_2\xi_3}^{(0)}
\Gamma_{\xi_2\zeta_3,\xi_4\gamma_1}^{(0)},
\end{eqnarray}
\begin{eqnarray}
\Sigma_{\zeta_1\zeta_2}^{\rm A3c\dag}(k) 
&= & \Bigl(\frac{T}{N}\Bigr)^3
\sum_{k,k_1,k_2}\sum_{\zeta_3\zeta_4}
\sum_{\xi_1\xi_2\xi_3\xi_4}
\sum_{\gamma_1\gamma_2\gamma_3\gamma_4}
F_{\zeta_4\zeta_3}^{\dag}(k')
G_{\gamma_3\gamma_4}^{(0)}(k-k'+k_1)
G_{\gamma_1\gamma_2}^{(0)}(k_1)\nonumber\\
&&\times G_{\xi_3\xi_2}^{(0)}(k+k_1+k_2)G_{\xi_1\xi_4}^{(0)}(k_2)
\Gamma_{\xi_4\gamma_2,\xi_3\zeta_1}^{(0)}
\Gamma_{\gamma_4\zeta_3,\zeta_2\gamma_1}^{(0)}
\Gamma_{\xi_2\zeta_4,\xi_1\gamma_3}^{(0)},
\end{eqnarray}
\begin{eqnarray}
\Sigma_{\zeta_1\zeta_2}^{\rm A3d\dag}(k) 
&= & \frac{1}{2}\Bigl(\frac{T}{N}\Bigr)^3
\sum_{k,k_1,k_2}\sum_{\zeta_3\zeta_4}
\sum_{\xi_1\xi_2\xi_3\xi_4}
\sum_{\gamma_1\gamma_2\gamma_3\gamma_4}
F_{\zeta_4\zeta_3}^{\dag}(k')
G_{\gamma_2\gamma_1}^{(0)}(k-k'+k_1)G_{\gamma_4\gamma_3}^{(0)}(k_1)\nonumber\\
&&\times G_{\xi_2\xi_3}^{(0)}(k+k_1-k_2)G_{\xi_4\xi_1}^{(0)}(k_2)
\Gamma_{\zeta_4\gamma_3,\gamma_2\zeta_1}^{(0)}
\Gamma_{\gamma_1\zeta_3,\xi_4\xi_2}^{(0)}
\Gamma_{\xi_1\xi_3,\gamma_4\zeta_2}^{(0)},
\end{eqnarray}
\begin{eqnarray}
\Sigma_{\zeta_1\zeta_2}^{\rm A3e\dag}(k) 
&= & \frac{1}{2}\Bigl(\frac{T}{N}\Bigr)^3
\sum_{k,k_1,k_2}\sum_{\zeta_3\zeta_4}
\sum_{\xi_1\xi_2\xi_3\xi_4}
\sum_{\gamma_1\gamma_2\gamma_3\gamma_4}
F_{\zeta_4\zeta_3}^{\dag}(k')\nonumber
G_{\gamma_4\gamma_3}^{(0)}(-k+k'+k_1)
G_{\gamma_2\gamma_1}^{(0)}(k_1)\nonumber\\
&&\times G_{\xi_2\xi_4}^{(0)}(-k+k_1-k_2)G_{\xi_1\xi_3}^{(0)}(k_2)
\Gamma_{\xi_3\xi_4,\gamma_2\zeta_1}^{(0)}
\Gamma_{\zeta_4\gamma_3,\xi_2\xi_1}^{(0)}
\Gamma_{\gamma_1\zeta_3,\zeta_2\gamma_4}^{(0)},
\end{eqnarray}

where $F_{\zeta_4\zeta_3}^{\dag}(k)$
is the anomalous Green's function.
The factors $\frac{1}{2}$ multiplied in the (A1), (A3d) and (A3e)
are necessary to avoid the redundancy in the summation. 
By summing up these contributions, 
we obtain the following \'Eliashberg equation 
within the third order perturbation theory. 
\begin{equation}
\Sigma_{\zeta_1\zeta_2}^{\rm A\dag}(k) = 
- \frac{T}{N}\sum_{k'}\sum_{\zeta_3\zeta_4}F_{\zeta_4\zeta_3}^{\dag}(k')
\bigl(\Gamma_{\zeta_4\zeta_3,\zeta_2\zeta_1}(k', k)
-\frac{1}{2}\Gamma^{(0)}_{\zeta_4\zeta_3,\zeta_2\zeta_1}\bigr), 
\label{Eq:Eleq}
\end{equation}
where $\Gamma_{\zeta_4\zeta_3,\zeta_2\zeta_1}(k', k)$ is 
the renormalized pair scattering amplitude and given in the following, 
\begin{eqnarray}
\Gamma_{\zeta_4\zeta_3,\zeta_2\zeta_1}(k', k) &=& 
\Gamma^{(0)}_{\zeta_4\zeta_3,\zeta_2\zeta_1}
-\sum_{\gamma_1\gamma_3}X^{(0)}_{{m_1}{m_3}}(k-k')
\Gamma^{(0)}_{\zeta_4\gamma_3,\gamma_1\zeta_1}
\Gamma^{(0)}_{\gamma_1\zeta_3,\zeta_2\gamma_3}\nonumber\\
&+&\sum_{\gamma_1\gamma_2}\sum_{\xi_1\xi_2}
X^{(0)}_{{n_1}{n_2}}(k-k')X^{(0)}_{{m_1}{m_2}}(k-k')
\Gamma^{(0)}_{\zeta_4\xi_2,\xi_1\zeta_1}
\Gamma^{(0)}_{\xi_1\gamma_2,\gamma_1\xi_2}
\Gamma^{(0)}_{\gamma_1\zeta_3,\zeta_2\gamma_2}\nonumber\\
&-&\frac{T}{N}\sum_{k_1}\sum_{\gamma_1\gamma_3}\sum_{\xi_1\xi_2}
G^{(0)}_{m_3}(-k+k'+k_1)
\Bigl[X^{(0)}_{{n_2}{n_1}}(-k+k_1)
\Gamma^{(0)}_{\zeta_4\gamma_3,\gamma_1\zeta_1}
\Gamma^{(0)}_{\gamma_1\xi_1,\zeta_2\xi_2}
\Gamma^{(0)}_{\xi_2\zeta_3,\xi_1\gamma_3}\nonumber\\
&&+\frac{1}{2}\Phi^{(0)}_{{n_2}{n_1}}(-k+k_1)
\Gamma^{(0)}_{\xi_2\xi_1,\zeta_1\gamma_1}
\Gamma^{(0)}_{\zeta_4\gamma_3,\xi_2\xi_1}
\Gamma^{(0)}_{\gamma_1\zeta_3,\zeta_2\gamma_3}
\Bigr]G^{(0)}_{m_1}(k_1)\nonumber\\
&-&\frac{T}{N}\sum_{k_1}\sum_{\gamma_1\gamma_3}
\sum_{\xi_1\xi_2}
G^{(0)}_{m_3}(k-k'+k_1)
\Bigl[X^{(0)}_{{n_2}{n_1}}(k+k_1)
\Gamma^{(0)}_{\xi_1\gamma_1,\xi_2\zeta_1}
\Gamma^{(0)}_{\gamma_3\zeta_3,\zeta_2\gamma_1}
\Gamma^{(0)}_{\xi_2\zeta_4,\xi_1\gamma_3}\nonumber\\
&&+\frac{1}{2}\Phi^{(0)}_{{n_2}{n_1}}(k+k_1)
\Gamma^{(0)}_{\zeta_4\gamma_1,\gamma_3\zeta_1}
\Gamma^{(0)}_{\gamma_3\zeta_3,\xi_1\xi_2}
\Gamma^{(0)}_{\xi_1\xi_2,\gamma_1\zeta_2}
\Bigr]G^{(0)}_{m_1}(k_1).
\label{Eq:Efint}
\end{eqnarray}
Here we have adopted the short notations, 
$\zeta_i=(\ell_i\sigma_i)$, $\gamma_i=(m_i\nu_i)$ 
and $\xi_i=(n_i\tau_i)$.
We assume the pairing interaction binds the most strongly 
the quasi-particles on the same band, 
and consider the three components of the anomalous Green's function 
and the anomalous self-energy, $F_{\nu\sigma\sigma'}^{\dag}(k)$ and 
$\Sigma_{\nu\sigma\sigma'}^{\rm A\dag}(k)$, where 
$\nu = \alpha, \beta, \gamma$. 
The diagonalized band indices $\nu$(=$\alpha$,$\beta$,$\gamma$) 
are replaced by the orbital indices $\ell$(=$xy$,$yz$,$xz$) 
through the following relations, 
\begin{eqnarray}
F_{\zeta\zeta'}^{\dag}(k)&=& \sum_{\nu=\alpha,\beta,\gamma}
U^{(0)}_{\ell\nu}(\mib{k})U^{(0)}_{\ell'\nu}(\mib{k})
F_{\nu\sigma\sigma'}^{\dag}(k)\\
&=& \sum_{\nu=\alpha,\beta,\gamma}
U^{(0)}_{\ell\nu}(\mib{k})U^{(0)}_{\ell'\nu}(\mib{k})|G_{\nu}(k)|^2
\Sigma_{\nu\sigma\sigma'}^{\rm A\dag}(k)\nonumber\\
\Sigma_{\nu\sigma\sigma'}^{\rm A\dag}(k)&=&\sum_{\ell\ell'}
{U^{(0)}_{\nu\ell}}^{-1}(\mib{k}){U^{(0)}_{\nu\ell'}}^{-1}(\mib{k})
\Sigma_{\zeta\zeta'}^{\rm A\dag}(k), 
\label{Eq:USig}
\end{eqnarray}
where $\zeta=(\ell \sigma)$, $\zeta'=(\ell' \sigma')$, 
and $\hat{U}^{(0)}({\mib k})$ is a matrix for the diagonalization 
of the noninteracting Hamiltonian $H_0$, 
\begin{equation}
\hat{U}^{(0)}(\mib{k})=
\begin{array}{r@{}l}
& \begin{array}{ccc}
\makebox[0.6em]{}\gamma &\makebox[1.1em]{}\alpha &\makebox[1.9em]{}\beta
\end{array} \\
\begin{array}{l}
{\it xy}\\ {\it yz}\\ {\it xz}\\
\end{array} & \left[
\begin{array}{ccc}
1 & 0 & 0 \\
0 & K(\mib{k}) & L(\mib{k})\\
0 & -L(\mib{k}) & K(\mib{k})
\end{array} \right],
\end{array}
\end{equation}
where 
\begin{eqnarray}
K(\mib{k}) &= &\sqrt{\frac{1}{2}(1-M(\mib{k}))},\\
L(\mib{k}) &= &{\rm sgn}(\lambda(\mib{k}))
\sqrt{\frac{1}{2}(1+M(\mib{k}))},\\
M(\mib{k}) &= &\frac{\xi_{-}(\mib{k})}{\sqrt{\xi_{-}^2(\mib{k})
+\lambda^2(\mib{k})}},\\
\xi_-(\mib{k}) &= &\frac{1}{2}(\xi_{yz}(\mib{k})-\xi_{xz}(\mib{k})). 
\end{eqnarray}
We replace the left hand side of the eq.~(\ref{Eq:Eleq}) 
by $\lambda(T)\Sigma_{\zeta_1\zeta_2}^{\rm A\dag}(q)$, 
and determine the transition temperature, at which the 
eigenvalue $\lambda(T)$ equals unity ($\lambda(T_{\rm c})=1.00$), 
by solving the set of the equations eqs.~(\ref{Eq:Eleq})-(\ref{Eq:USig}).
It is convenient to express the anomalous self-energy 
in the following form~\cite{Rf:Nomu.2}. 
For triplet pair, 
\begin{equation}
\Sigma_{\nu\sigma\sigma'}^{\rm A}(k)
=[{\rm i}(\mib{D}_\nu(k) \cdot \mib{\sigma})\sigma_y]_{\sigma\sigma'}, 
\end{equation}
and for singlet pair, 
\begin{equation}
\Sigma_{\nu\sigma\sigma'}^{\rm A}(k)
=[{\rm i}\Psi_\nu(k)\sigma_y]_{\sigma\sigma'}, 
\end{equation}
where $\mib{\sigma}$ is the Pauli matrix, 
and the function, $\mib{D}_\nu(k)$ ($\Psi_\nu(k)$), 
behaves as a vector(scalar) under the rotation in the spin space.
In the numerical calculation, we assume the vector $\mib{D}_\nu(k)$ 
is perpendicular to the basal plane, 
\begin{equation}
\mib{D}_\nu(k)=D_\nu(k)\hat{z}.
\end{equation}
The direction is consistent with the experimental results 
of Knight shift~\cite{Rf:Ishi.1}. 
This assumption is not substantial in the present calculation, 
because we do not introduce any anisotropy in the spin space. 
We can show that, if we take other direction of the vector, 
we obtain the same transition temperature. 
In order to determine the direction of the vector $\mib{D}_\nu(k)$, 
we must take into account the effect 
of the spin-orbit coupling~\cite{Rf:Ng.1}. 
We do not consider that the spin-orbit coupling is essential 
for discussing a mechanism of the spin-triplet superconductivity 
in Sr$_2$RuO$_4$. 

Here we define the function, 
$\Gamma_{\nu'\sigma_4\sigma_3,\nu\sigma_2\sigma_1}(k', k)$, by 
\begin{equation}
\Gamma_{\nu'\sigma_4\sigma_3,\nu\sigma_2\sigma_1}(k', k)
= \sum_{\ell_1\ell_2\ell_3\ell_4}
U^{(0)}_{\ell_4\nu'}(\mib{k}')U^{(0)}_{\ell_3\nu'}(\mib{k}')
\Gamma_{\zeta_4\zeta_3,\zeta_2\zeta_1}(k', k)
{U^{(0)}_{\nu\ell_2}}^{-1}(\mib{k}){U^{(0)}_{\nu\ell_1}}^{-1}(\mib{k}). 
\end{equation}
As we see in \S~\ref{Sc:Effint}, 
the component for the parallel spin pair 
on the $\gamma$ band, 
$\Gamma_{\gamma\sigma\sigma,\gamma\sigma\sigma}(k', k)$, 
possesses a momentum dependence favoring $p$-wave pairing. 

In conclusion of this section, 
note that the normal self-energy and 
the effective interaction have been 
expanded up to the third order in Coulomb energy 
by using only the bare $G^{(0)}_{\ell}(k)$ ($\ell = \{xy, yz, xz\}$), 
but the quasi-particles forming Cooper pair are described 
by the renormalized $G_{\nu}(k)$ $\nu = \{\alpha, \beta, \gamma \}$. 

\section{Results of Calculation}
\label{Sc:Results}

\subsection{Details of numerical calculations}

The momentum and frequency summations appearing in the 
equations, (\ref{Eq:N3})-(\ref{Eq:Phi0}), 
(\ref{Eq:Eleq}) and (\ref{Eq:Efint}) can be performed 
numerically with use of fast Fourier transformation algorithm. 
For the numerical calculations, the first Brillouin 
zone is divided into 128 $\times$ 128 $\mib{k}$-meshes, 
and the number of Matsubara frequencies taken is $N_f=1024$. 
The temperature region for numerical calculations is bounded from below 
by $T > W/ 2\pi N_f \approx 0.0012 $, 
where $W$ is the noninteracting band width. 
All of the numerical calculations in the present paper are 
performed in $T \geq 0.00250$. 
This lower bound corresponds to about 10(K), 
if we assume the band width $W$ is about 2(eV), 
and is still higher than the real $T_{\rm c}$ 
of Sr$_2$RuO$_4$, 1.5(K). 
Therefore we must extrapolate the calculation results 
of $T_{\rm c}$ to the weak interaction region. 

\subsection{Normal Fermi liquid state properties}
\label{Sc:Normal}

We show here the results of the quantities 
related to the normal Fermi liquid state. 

The Fermi surface consists of three sheets 
as depicted in Fig.~\ref{Fg:FermiS}. 
The circle around the point $(\pi, \pi)$, $\alpha$, is hole-like, 
while the large circles around the corners, $\beta$ and $\gamma$, 
enclose electrons. 
These results are in good agreement with those obtained by 
de Haas-van Alphen measurements~\cite{Rf:Mack.2}
and recent angle-resolved-photoemission 
measurements~\cite{Rf:Dama.1}. 

The normal self-energy on the $\gamma$ band as a function of frequency, 
$\Sigma_{\gamma}(\mib{k}_{\rm F},\omega)$, is shown in Fig.~\ref{Fg:self}. 
We can see usual Fermi liquid behaviors 
for every case of various inter-orbit couplings, 
${\rm Re} \Sigma(\omega) \sim c \omega $ 
and ${\rm Im} \Sigma(\omega) \sim c' \omega^2 $ ($c,c'< 0$) 
at low energy ($\omega \sim 0$). 
The remarkable property which we would like to point out here 
is that inter-orbit couplings do not affect 
the normal self-energy in the low energy region $\omega \approx 0$, 
while they do in the high energy region. 
This is partly due to the situation where the hybridization among 
the localized orbits, $xy$, $yz$ and $xz$, is not so large in Sr$_2$RuO$_4$. 
As far as we discuss low energy phenomena, such as 
superconductivity, therefore, we may expect that 
the single-band approach~\cite{Rf:Nomu.1} is a good starting point. 

Next we show the density of states(DOS) 
for a typical case in Fig.~\ref{Fg:dos}. 
The DOS of each band $\nu$ (= $\alpha$, $\beta$, $\gamma$) 
is calculated by the formula, 
\begin{equation}
\rho_{\nu}(\omega)=-\frac{1}{\pi}\sum_{\mib{k}}
{\rm Im} G^{\rm R}_{\nu}(\mib{k}, \omega), 
\end{equation} 
where $G^{\rm R}_{\nu}(k)$ is obtained by continuing analytically 
$G_{\nu}(k)$ from the upper half plane. 
What we should note is the large DOS peak near the Fermi level. 
This peak originates from the van Hove singularity 
on the two-dimensional band $\gamma$. 
As pointed out by Singh~\cite{Rf:Sing.1}, 
if the Fermi level comes above 
the singularity, the Fermi surface $\gamma$ changes its topology, 
from electron-like circle 
to hole-like circle around the point ($\pi$,$\pi$). 
The $\gamma$ band takes the main part of the DOS at the Fermi level. 
This agrees with the results 
of the de Haas-van Alphen measurements~\cite{Rf:Mack.2}. 
As a result, the $\gamma$ band is expected to be significant 
in discussing the electronic properties of Sr$_2$RuO$_4$. 

\subsection{Eigenvalues for triplet $p$-wave and singlet $d$-wave states}
\label{Sc:Egnvls}

We show the calculation results of the maximum eigenvalues 
for the spin-triplet and the singlet states 
in Fig.~\ref{Fg:pdegnvls}. There we show the calculation results 
for four cases of relatively strong inter-orbit couplings. 
The orbital symmetry of the state giving the maximum eigenvalues 
is $p$-wave for the spin-triplet state and $d_{x^2-y^2}$-wave 
for the spin-singlet state. 
In the cases (a), (b) and (d), 
we obtain the larger eigenvalues for the spin-triplet state 
than for the spin-singlet state at low temperatures. 
Therefore we expect that the $p$-wave state is realized 
for the three cases. 
For the case (c), where the Hund's coupling 
is strong, we can see that the eigenvalues for the $p$-wave state 
is small compared with those for the $d_{x^2-y^2}$ state. 
The reason why the $d_{x^2-y^2}$-wave pairing state 
becomes more stable than $p$-wave one, as the strength of 
the inter-orbit coupling increases, will be discussed 
in \S~\ref{Sc:DiscConc}. 
It should be noted here that the values of inter-orbit couplings  
in the cases, (a), (b) and (c) are too large. 
We consider that actually they should be smaller, 
since in general they are weakened more 
by the screening effects than the intra-orbit coupling $U$. 
If we take smaller realistic inter-orbit couplings, 
for example, as in the case (d), 
$d_{x^2-y^2}$-wave state is not expected to be realized at all. 
Note that the situations for which we have given the results 
of eigenvalues here are still relatively unfavorable for the $p$-wave state 
because of the strong inter-orbit couplings. 

\subsection{Transition temperature for triplet $p$-wave state}
\label{Sc:T_c}

The calculated transition temperature for the $p$-wave state 
is displayed in Fig.~\ref{Fg:T_c} as a function of $U$. 
In the both cases of the inter-orbit couplings, (a) and (b), 
we may obtain higher transition temperature for the $p$-wave state, 
according to the results of calculation of eigenvalues 
in \S~\ref{Sc:Egnvls}. 
We consider that the lower bound of temperature 
for the calculation, $T=0.00300$, 
corresponds to about 10(K). 
We may obtain the more realistic value of $T_{\rm c}$=1.5(K), 
by extrapolating the results down to the smaller $U$ region. 
According to the results, $T_{\rm c}$ becomes high 
for strong inter-orbit couplings. 
In the limit of weak inter-orbit coupling, 
we obtain the same results as we did in the previous 
single-band analysis for the $\gamma$ band. 

\subsection{Momentum dependence of anomalous self-energy parts}
\label{Sc:Anomalous}

The momentum dependence of the anomalous self-energy 
$D_{\nu}(k)$ is plotted in the Fig.~\ref{Fg:DVq}, 
where we consider the state with the $k_y$-like symmetry 
$D_{\nu}(k_x, k_y, {\rm i}\omega_n)= -D_{\nu}(k_x, -k_y, {\rm i}\omega_n)$. 
The results show highly anisotropic $p$-wave state, 
and not $f$-wave. 
If we assume in the beginning the condition, 
$D_{\nu}(k_x, k_y, {\rm i}\omega_n)
= -D_{\nu}(-k_x, k_y, {\rm i}\omega_n)$, 
we obtain the $p$-wave state with the $k_x$-like symmetry. 
The relative phases of $D_{\nu}(k)$'s converge to zero, 
and we can assume without any loss of the generality 
that $D_{\nu}(k)$'s are real. 

We should note that the magnitude of the anomalous self-energy 
is the largest on the $\gamma$ Fermi surface. 
This is ``orbital dependent superconductivity", 
which was proposed originally 
by Agterberg {\it et al}~\cite{Rf:Agte.1}. 
The magnitudes of $D_{\alpha}(k)$ and $D_{\beta}(k)$ 
are almost the same, since both 
of the $\alpha$ and the $\beta$ bands 
are constructed by hybridizing the original 
one-dimensional $yz$ and $xz$ bands. 
The $\gamma$ band originates mainly 
from the two-dimensional $xy$ band, 
and is separated from the other two bands, 
$\alpha$ and $\beta$, in that 
the hybridization is expected to be small 
(In the present theoretical formulation, 
the hybridization is assumed to be exactly zero, 
since we consider only the single RuO$_2$ layer). 
In Fig.~\ref{Fg:DVq}, we have shown the case of 
relatively strong inter-orbit couplings. 
If we take small inter-orbit couplings, 
the relative magnitude of $D_{\gamma}(k)$ 
to $D_{\alpha,\beta}(k)$ becomes large. 
In the unrealistic and extreme case where inter-orbit couplings are 
zero ($U'=J=J'=0$), $D_{\alpha}(k)$ and $D_{\beta}(k)$ 
converge to zero, 
and only $D_{\gamma}(k)$ converges to a finite value. 
As a consequence, we find that the $\gamma$ band 
dominates the superconducting transition. 
Here we have shown only the results for the triplet 
$p$-wave state. According to our calculation, 
the dominance of the $\gamma$ band is also valid for 
the case of the singlet $d_{x^2-y^2}$-wave state. 

\subsection{Momentum dependence of effective pairing interaction 
on the $\gamma$ band}
\label{Sc:Effint}

We extract only the pair scattering amplitude of quasi-particles 
from the $\gamma$ band to the $\gamma$ band. 
This channel is the most significant not only for the $p$-wave 
pairing but also for the $d_{x^2-y^2}$-wave pairing. 
In Fig.~\ref{Fg:effint} we show the plots of the vertex functions 
(a)$\Gamma_{\gamma\sigma\sigma,\gamma\sigma\sigma}(k', k)$ 
for the parallel spin pairs and 
(b)$\Gamma_{\gamma\sigma\bar{\sigma},\gamma\bar{\sigma}\sigma}(k', k)$
for the anti-parallel spin pairs. 

In the triplet case (Fig.~\ref{Fg:effint}(a)), 
we find that the function 
$\Gamma_{\gamma\sigma\sigma,\gamma\sigma\sigma}(k', k)$, 
has a maximum value around the point $\mib{k} \approx -\mib{k}'$. 
This feature originates from the vertex correction terms
(mainly, (A3d) and (A3e) in Fig.~\ref{Fg:aself}), 
and not from the exchange of any bosonic excitations. 
This characteristic momentum dependence of the vertex function is 
the origin of the relatively high transition temperature 
for the triplet pairing state, and is basically 
the same one as derived in the previous work 
for the single-band Hubbard model~\cite{Rf:Nomu.1}. 

In the singlet case (Fig.\ref{Fg:effint}(b)), 
$\Gamma_{\gamma\sigma\bar{\sigma},\gamma\bar{\sigma}\sigma}(q; k)$ 
shows the characteristic peak around 
$\mib{k}=\mib{k}'+\mib{Q}_{\rm IAF}$, 
where $\mib{Q}_{\rm IAF} \approx (\pm 0.6 \pi, \pm 0.6 \pi)$. 
These peaks are due to the nesting of the 
quasi-one-dimensional Fermi surfaces, $yz$ and $xz$ 
(or almost equivalently, $\alpha$ and $\beta$). 
This momentum dependence was observed as 
sizeable incommensurate antiferromagnetic spin fluctuations 
by the inelastic neutron scattering measurements~\cite{Rf:Sidi.1}, 
and was obtained also by the theoretical calculations 
on the magnetic properties of Sr$_2$RuO$_4$~\cite{Rf:Mazi.2, Rf:Nomu.3}. 
Since it is through the inter-orbit couplings that 
the momentum dependence of the incommensurate fluctuations 
is reflected in the pairing interaction on the $\gamma$ band, 
the peaks around $\mib{k}=\mib{k}'+\mib{Q}_{\rm IAF}$ vanish 
as the strength of the inter-orbit couplings is decreased. 
On the other hand, the large values of the vertex function 
around $\mib{k} \approx \mib{k}'$ are due to the 
ferromagnetic components of the susceptibility 
for the $xy$($\gamma$) band. 
Note that the values of the vertex function 
$\Gamma_{\gamma\sigma\bar{\sigma},\gamma\bar{\sigma}\sigma}(k', k)$ 
are positive all over the Brillouin zone. 
This is because of the on-site repulsion $U$. 
In general this is the reason why conventional $s$-wave superconductivity 
is not expected to be realized in strongly correlated electron systems.  

Note that, as mentioned in \S~\ref{Sc:Egnvls}, 
the cases we have shown here are the strong inter-orbit coupling ones. 
We consider that the inter-orbit couplings are actually not so strong. 
We have assumed such strong inter-orbit couplings 
to elucidate their roles. 

\section{Discussion and Conclusion}
\label{Sc:DiscConc}

In this section, 
we would like to discuss the results of calculation comprehensively, 
and give some proposals on experiments, based on our microscopic 
picture of the superconductivity in Sr$_2$RuO$_4$. 

In \S~\ref{Sc:Model}, we have taken the electron filling 
per one spin state, $n_{xy}=n_{yz}=n_{xz}=0.700$. 
This seems to be a little large compared with 
the results of the de Haas-van Alphen measurements~\cite{Rf:Mack.2}, 
although the Fermi surfaces for this filling 
reproduce the observed ones qualitatively well, as shown 
in \S~\ref{Sc:Normal}. 
According to the measurements, 
it is plausible that the orbits are filled 
as the total electron number equals 4.032. 
If we assume that all of the three orbits are filled 
with even number of electrons, the electron number 
of one orbit is $n_{\ell}=0.672$.
If we take the filling $n_{\ell}=0.672$ in the calculation, 
the parameter region of the inter-orbit couplings 
where the $p$-wave state overcomes the $d_{x^2-y^2}$-wave state 
is limited only in the weak inter-orbit couplings. 
We consider, however, that this situation is unrealistic, 
because $d_{x^2-y^2}$-wave state has never been observed 
experimentally near the triplet superconducting phase of Sr$_2$RuO$_4$. 
We relax the relation $2(n_{xy}+n_{yz}+n_{xz})=4.032$ 
in order to obtain the good agreement with the realistic situation. 
The incommensurate nesting vector $\mib{Q}_{\rm IAF}$ 
(=(${Q_{\rm IAF}}_x, {Q_{\rm IAF}}_y$, 0)) is 
estimated by ${Q_{\rm IAF}}_y \approx 2(1-n_{yz})\pi$ 
and ${Q_{\rm IAF}}_x \approx 2(1-n_{xz})\pi$. 
The observed nesting vector is ${Q_{\rm IAF}}_x
={Q_{\rm IAF}}_y=0.60\pi$~\cite{Rf:Sidi.1}. 
Therefore we should take $n_{yz}=n_{xz}=0.700$, 
and we have assumed that the three orbits are 
almost degenerate and $n_{xy}=0.700$. 
At the present stage, the theory proposed here suggests 
that Sr$_2$RuO$_4$ seems to be located near the boundary 
of the $p$-wave and the $d_{x^2-y^2}$ states in the parameter space. 
We consider, however,  that the present theory may be still insufficient 
for discussing quantitatively the competition of the $p$-wave 
and the $d_{x^2-y^2}$-wave pairing states. 
In the future, it must be proved by more refined theoretical treatments 
that the $d_{x^2-y^2}$-wave pairing state is more suppressed. 
One way of the further study is to inspect the effect of the higer order terms. 
In any case, we believe that, as far as we discuss only the mechanism 
of the $p$-wave superconductivity in Sr$_2$RuO$_4$, 
the theory gives a satisfactory result, 
in that we have obtained the relatively high $T_{\rm c}$ 
for the $p$-wave state and the momentum dependence which 
favors the pairing state, as shown 
in Figs.~\ref{Fg:T_c} and \ref{Fg:effint}(a). 

In \S~\ref{Sc:Egnvls} we have shown that 
for the strong inter-orbit couplings, 
particularly for the strong Hund's coupling, 
the eigenvalues for the $d_{x^2-y^2}$-wave state become large 
compared with those for the $p$-wave state. 
This is because the stronger the inter-orbit coupling becomes, 
the more prominently the incommensurate spin fluctuations 
are reflected in the pairing interaction on the $\gamma$ band. 
As the present authors pointed out in the previous work 
on the magnetic properties of quasi-two-dimensional ruthenates~\cite{Rf:Nomu.3}, 
the Hund's coupling enhances the incommensurate fluctuations rather 
than the ferromagnetic components of the spin susceptibility. 
Therefore we may conclude that the Hund's coupling disturbs 
the $p$-wave pairing by enhancing the $d$-wave pairing. 

As shown in \S~\ref{Sc:Anomalous}, 
the orbital symmetry of the pairing is anisotropic $p$-wave. 
In general, if we assume the most promising form of 
the vector $\mib{d}(\mib{k}) \sim (k_x \pm {\rm i}k_y)\hat{z}$~\cite{Rf:Sigr.1}, 
we usually expect the nodeless gap around the cylindrical Fermi surface. 
The nodeless energy gap in Sr$_2$RuO$_4$ has, however, 
been considered to be inconsistent with the observed power-law 
behaviors in various quantities 
at low temperatures~\cite{Rf:Ishi.2,Rf:Nish.1,Rf:Bona.1,Rf:Lupi.1}. 
Recently we have successfully shown that 
the $p$-wave superconducting gap derived in the present formulation 
is quite consistent with the power-law behavior 
of the specific heat below $T_{\rm c}$~\cite{Rf:Nomu.4}. 
There the strong momentum dependence 
of the anomalous self-energy is essential 
for a node-like structure on the $\beta$ Fermi surface, 
and results in the power-law behavior at the low temperature. 
Accordingly, we would like to point out here that 
the $p$-wave state obtained in our discussions 
can explain the power-law behaviors of the specific heat, 
even if we assume the symmetry 
$\mib{d}(\mib{k}) \sim (k_x \pm {\rm i}k_y)\hat{z}$. 

We would like to give some proposals for experiments, 
based on our microscopic picture on the superconductivity of Sr$_2$RuO$_4$. 
Let us consider that the number of electrons 
in the band $\gamma$($xy$) is increased. 
In such a case, as we previously showed 
in the single-band analysis for the $\gamma$ band~\cite{Rf:Nomu.1}, 
the superconducting transition temperature 
is expected to be enhanced. 
This is due to the situation where, if we increase the electrons 
in $\gamma$ band, the Fermi level becomes close to 
the van Hove singularity on the $\gamma$ band and 
the density of states at the Fermi level is increased. 
Note here that the van Hove singularity is located only 
slightly above the Fermi level, 
as shown in Fig.~\ref{Fg:dos}. 
At the same time, as we insisted in the previous work~\cite{Rf:Nomu.3}, 
the ferromagnetic components of the spin susceptibility 
may be enhanced. 
Carrier dopings by chemical substitution 
usually not only change the carrier number 
but also damage the conducting RuO$_2$ plane. 
Since the triplet superconductivity is sensitively 
destroyed by disorders, we should not take any 
chemical substitution for the carrier doping. 
We think that the most practical way of increasing 
the electron number in the $\gamma$ band 
without damaging the RuO$_2$ layers at all is application of 
the uniaxial pressure along the $c$-axis. 
The orbits $yz$ and $xz$ spatially extends along the $c$-axis, 
while the orbit $xy$ will be flat along the basal plane. 
If we apply the uniaxial pressure, the energy levels of 
the orbits $yz$ and $xz$ are lifted 
due to the crystalline field effect, 
while the $xy$ orbit is not. 
Therefore we may expect that the electrons are 
transferred from the $yz$ and the $xz$ orbits to the $xy$ orbit. 
Consequently, we would like to point out the following possibility. 
If we can indeed transfer the electrons from the $yz$ 
and the $xz$ orbits to the $xy$ orbit by applying the uniaxial pressure, 
the transition temperature may become higher. 
At the same time, the system may become close to the ferromagnetism, 
as we suggested in the previous work~\cite{Rf:Nomu.3}. 
Measuring the transition temperature and the uniform spin susceptibility  
under the uniaxial pressure along the $c$-axis is, therefore, very interesting. 
Recent experimental work on the elastic moduli suggests the possibility 
that the transition temperature is increased by applying the uniaxial pressure 
along the $c$-axis~\cite{Rf:Okuda.1}. 
We may consider the fact as an evidence that the superconducting transition 
occurs predominantly on the $\gamma$ band. 
This experimental fact supports the $p$-wave pairing rather than 
the $d$-wave pairing, because, if the $d_{x^2-y^2}$-wave state was realized, 
the transition temperature should become lower 
by increasing the electron number of the $\gamma$ band, 
according to our previous work~\cite{Rf:Nomu.1}. 

In conclusion, we summarize the main results obtained. 
We have discussed a mechanism 
of the spin-triplet superconductivity in Sr$_2$RuO$_4$. 
We can obtain the $p$-wave pairing state for moderately 
strong inter-orbit couplings. 
There one of the three bands, $\gamma$, 
plays the dominant role in the superconducting transition, 
and the pairing on the other two bands($\alpha$ and $\beta$) 
is induced passively through the inter-orbit couplings. 
The most significant momentum dependence for the $p$-wave pairing 
originates from the vertex correction terms, 
and is basically the same one as we obtained 
in the previous single-band analysis~\cite{Rf:Nomu.1}. 
Therefore we should regard the spin-triplet superconductivity 
in Sr$_2$RuO$_4$ as one of the natural results of electron correlations, 
and cannot consider as a result of some strong magnetic fluctuations. 
We would like to insist that Sr$_2$RuO$_4$ realizes 
the $p$-wave superconductivity which is basically described 
by a simple one-band repulsive Hubbard model~\cite{Rf:Nomu.1}. 

\section*{Acknowledgments}
The authors are grateful for the useful discussions 
to Professor Y. Maeno, Professor M. Sigrist, 
Professor K. Ishida and Dr. N. Kikugawa. 
They would like to thank Dr. N. Okuda and Professor T. Suzuki 
for the generosity. 
Numerical calculations in this work were performed 
at the Yukawa Institute Computer Facility.

\begin{figure}
\caption{Schematic illustration of single RuO$_2$ layer.
The Ru4d$_{xy, yz, xz}$ and O2p$_{x, y, z}$ orbitals 
construct the two-dimensional network 
by hybridizing with each other.} 
\label{Fg:model}

\caption{The antisymmetric bare vertex, 
$\Gamma^{(0)}_{\zeta_1\zeta_2,\zeta_3\zeta_4}$.}
\label{Fg:bvert}

\caption{
Perturbation terms for normal self-energy 
up to the third order. The shaded circle in (N1) 
represents arbitrary self-energy insertion. 
The solid line and the empty square 
denote the bare Green's function $G^{(0)}_{\ell}(k)$ 
and the bare antisymmetric vertex $\Gamma^{(0)}$, 
respectively.}
\label{Fg:nself}

\caption{
Perturbation terms for anomalous self-energy 
up to the third order.
The thin solid line, the thick solid line and the empty square 
denote the bare Green's function $G^{(0)}_{\ell}(k)$, the anomalous 
Green's function $F_{\zeta\zeta'}^{\dag}(k)$ 
and the antisymmetric bare vertex $\Gamma^{(0)}$, 
respectively.}
\label{Fg:aself}

\caption{The Fermi surface. The Coulomb integrals are $U=3.731$, 
$U'=0.300U$, $J=J'=U'$, and the temperature is $T=0.0100$.}
\label{Fg:FermiS}

\caption{
The normal self-energy on the $\gamma$ Fermi surface 
as a function of frequency. 
(a) real part, (b) imaginary part. }
\label{Fg:self}

\caption{The density of states. 
The Fermi level corresponds to $\omega=0$.
(a)The total density of states.
(b)The partial density of states for the three bands, 
$\alpha$, $\beta$ and $\gamma$.
The inset shows the details near the Fermi level.}
\label{Fg:dos}

\caption{
The maximum eigenvalues as a function of temperature $T$. 
For all of the cases, the intra-orbit coupling is fixed as $U=3.079$. 
(a) $U'=0.500U$, $J=J'=0.667U'$. 
(b) $U'=0.500U$, $J=J'=U'$.
(c) $U'=0.500U$, $J=J'=1.333U'$.
(d) $U'=0.300U$, $J=J'=U'$.
In the cases, (a), (b) and (d), 
we expect that the spin-triplet $p$-wave state 
is stable compared with the spin-singlet $d_{x^2-y^2}$ state.}
\label{Fg:pdegnvls}

\caption{
Transition temperature as a function of $U$ 
for spin-triplet $p$-wave state. 
(a) $J=J'=0.667U'$.
(b) $J=J'=U'$.}
\label{Fg:T_c}

\caption{
Contourplots of  the results of 
$D_{\nu}(\mib{k},{\rm i}\pi T)$ ($\nu$=$\alpha$, $\beta$, $\gamma$). 
In the light and the dark colored regions, 
the functions take higher and lower values, respectively. 
The thick line circles represent the Fermi surfaces. 
The parameters are $U=3.385$, $U'=0.500U$, $J=J'=U'$ and $T=0.00300$.}
\label{Fg:DVq}

\caption{
Contourplots of effective interaction on the $\gamma$ band. 
The thick line circle around the corner 
is the Fermi surface $\gamma$. 
(a) For the quasi-particles with parallel spins, 
$\Gamma_{\gamma\sigma\sigma,\gamma\sigma\sigma}
(\mib{k}',{\rm i}\pi T; \mib{k},{\rm i}\pi T)$. 
(b) For the quasi-particles with anti-parallel spins, 
$\Gamma_{\gamma\sigma\bar{\sigma},\gamma\bar{\sigma}\sigma}
(\mib{k}',{\rm i}\pi T; \mib{k},{\rm i}\pi T)$. 
In both cases, $\mib{k}'$ is fixed as pointed by the arrow. 
The parameters are $U=3.385$, $U'=0.500U$, $J=J'=U'$ and $T=0.00700$.}
\label{Fg:effint}
\end{figure}

\clearpage
\setcounter{figure}{0}

\clearpage
\begin{figure}
\leavevmode
\epsfxsize=100mm
\epsfbox{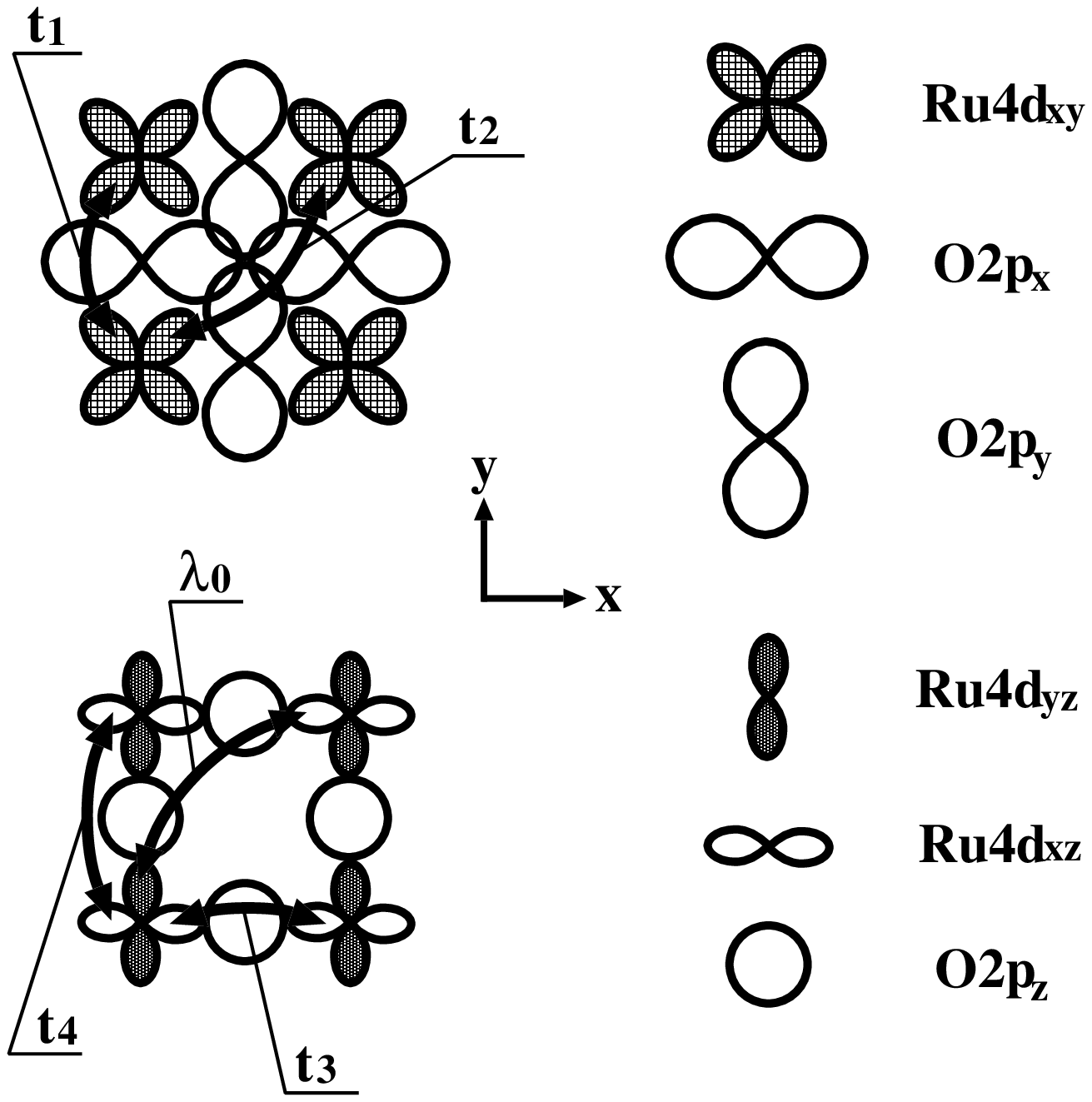}
\caption{T. Nomura \& K. Yamada}
\end{figure}

\clearpage
\begin{figure}
\leavevmode
\epsfxsize=80mm
\epsfbox{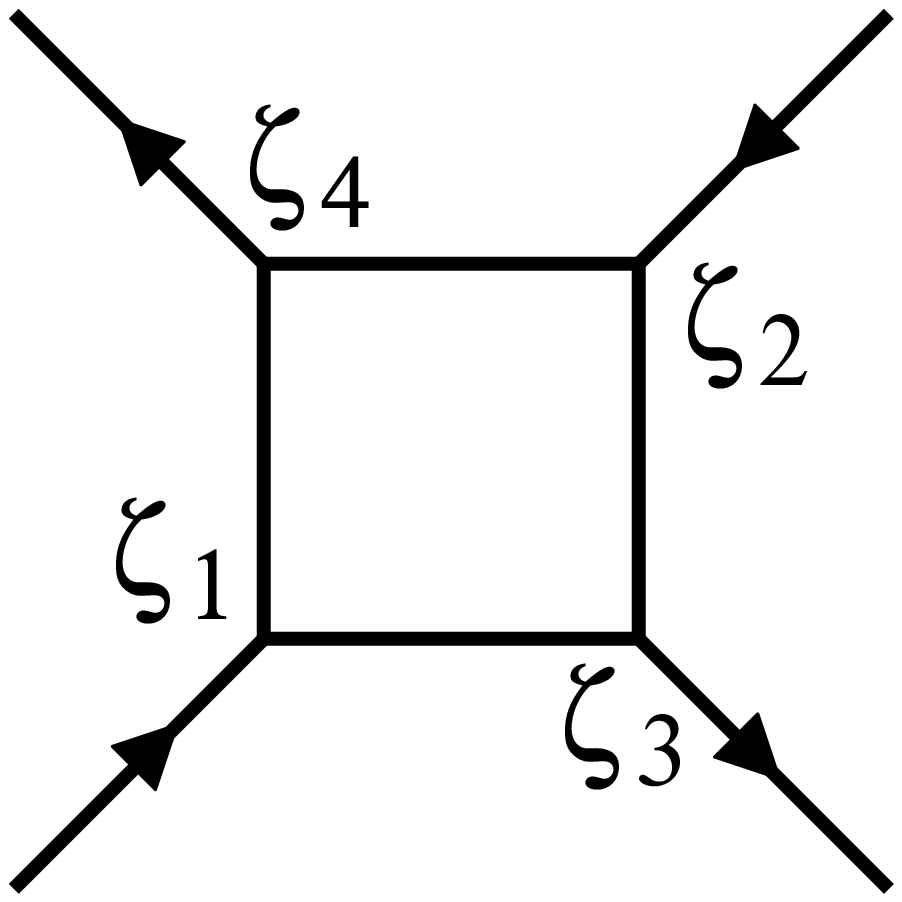}
\caption{T. Nomura \& K. Yamada}
\end{figure}

\clearpage
\begin{figure}
\leavevmode
\epsfxsize=120mm
\epsfbox{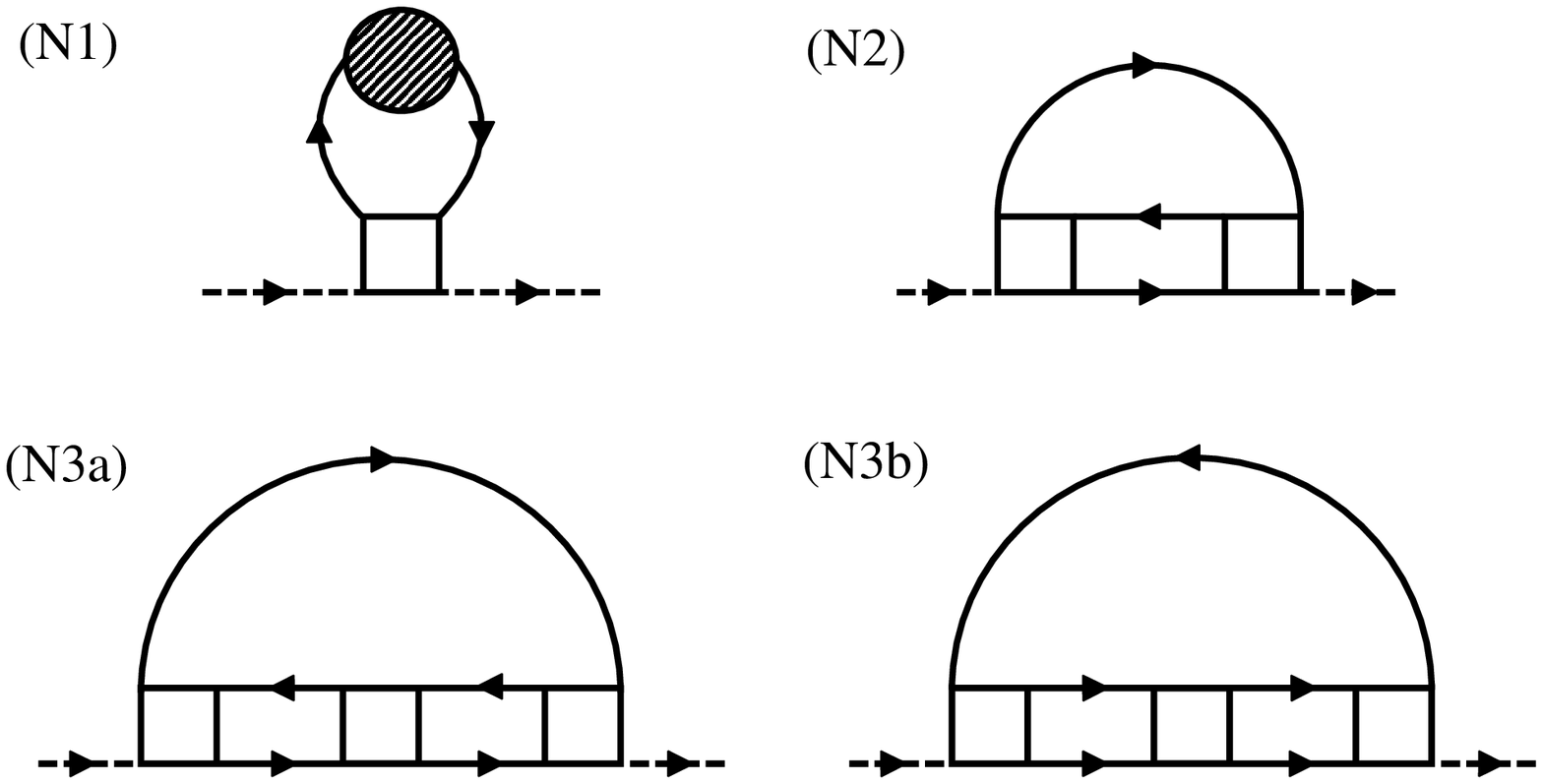}
\caption{T. Nomura \& K. Yamada}
\end{figure}

\clearpage
\begin{figure}
\leavevmode
\epsfxsize=120mm
\epsfbox{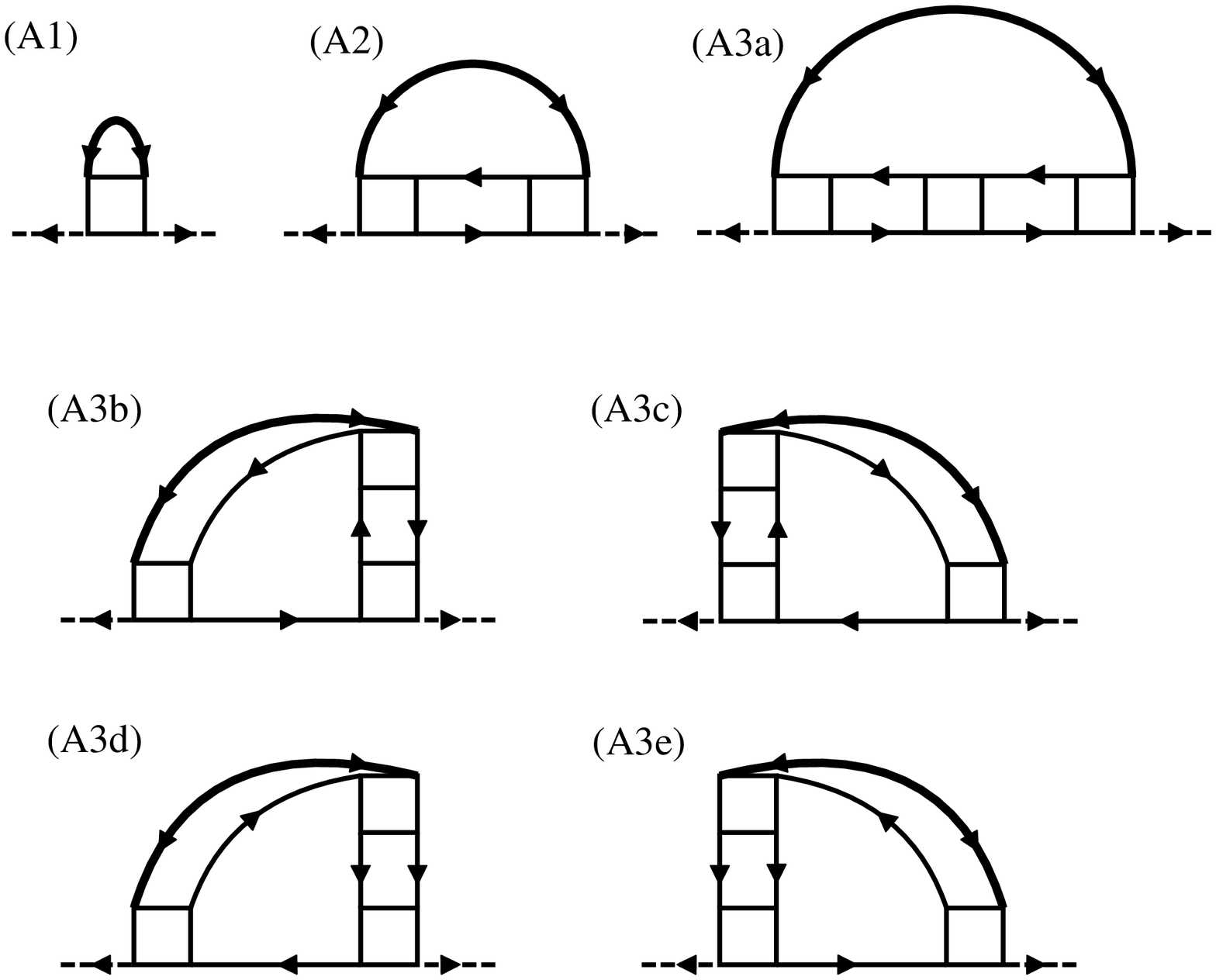}
\caption{T. Nomura \& K. Yamada}
\end{figure}

\clearpage
\begin{figure}
\leavevmode
\epsfxsize=120mm
\epsfbox{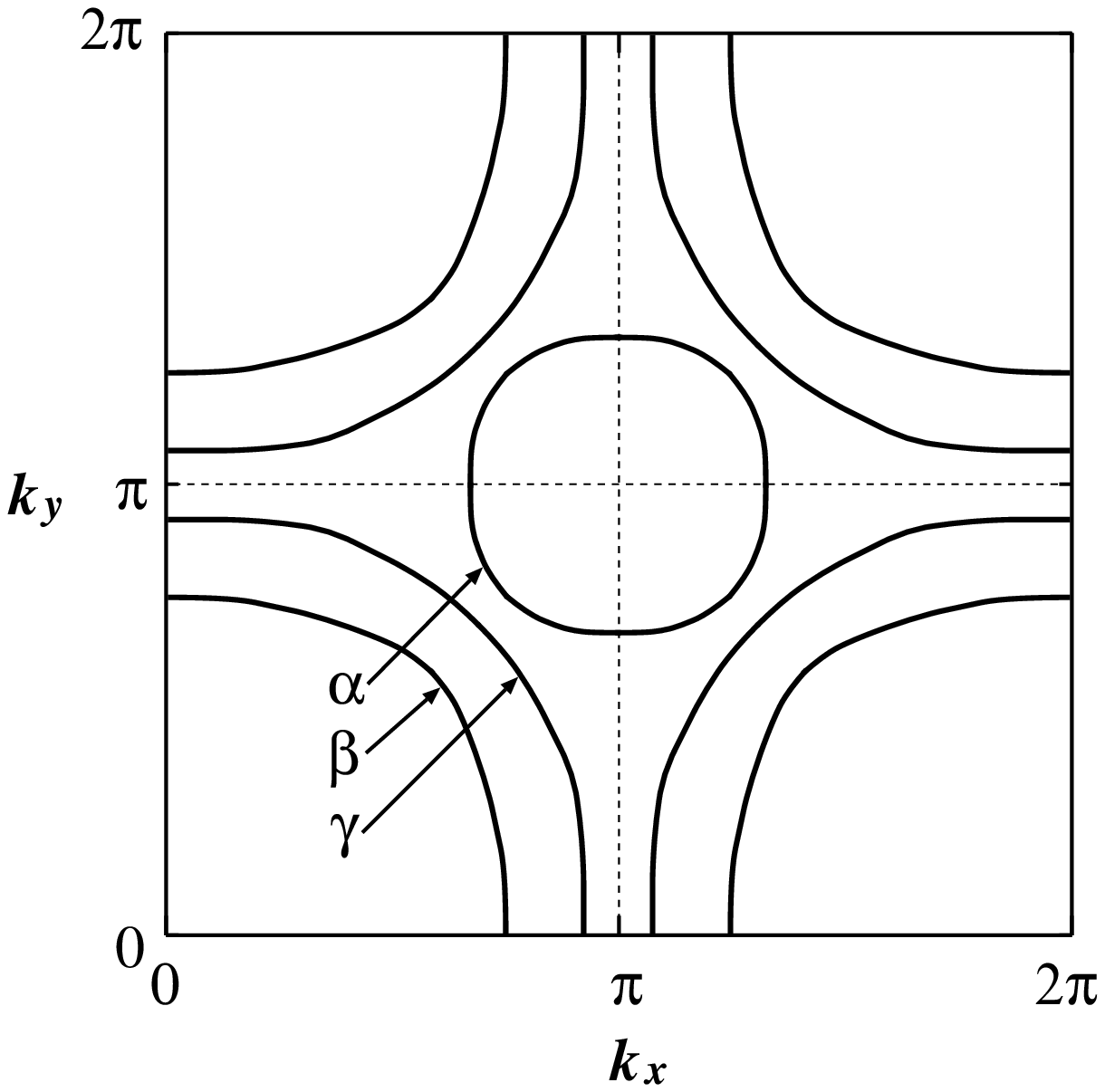}
\caption{T. Nomura \& K. Yamada}
\end{figure}

\clearpage
\begin{figure}
\leavevmode
\epsfxsize=120mm
\epsfbox{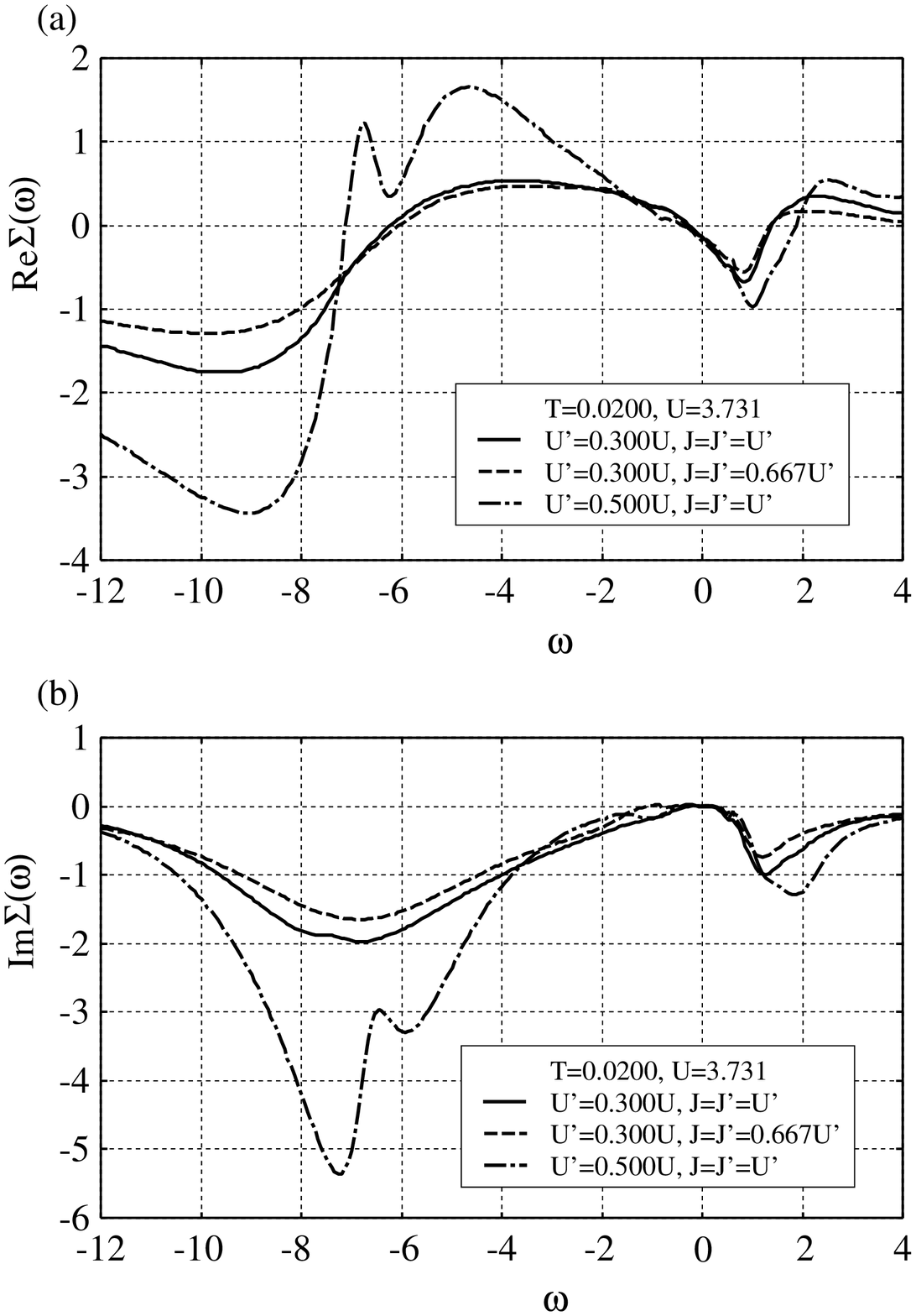}
\caption{T. Nomura \& K. Yamada}
\end{figure}

\clearpage
\begin{figure}
\leavevmode
\epsfxsize=120mm
\epsfbox{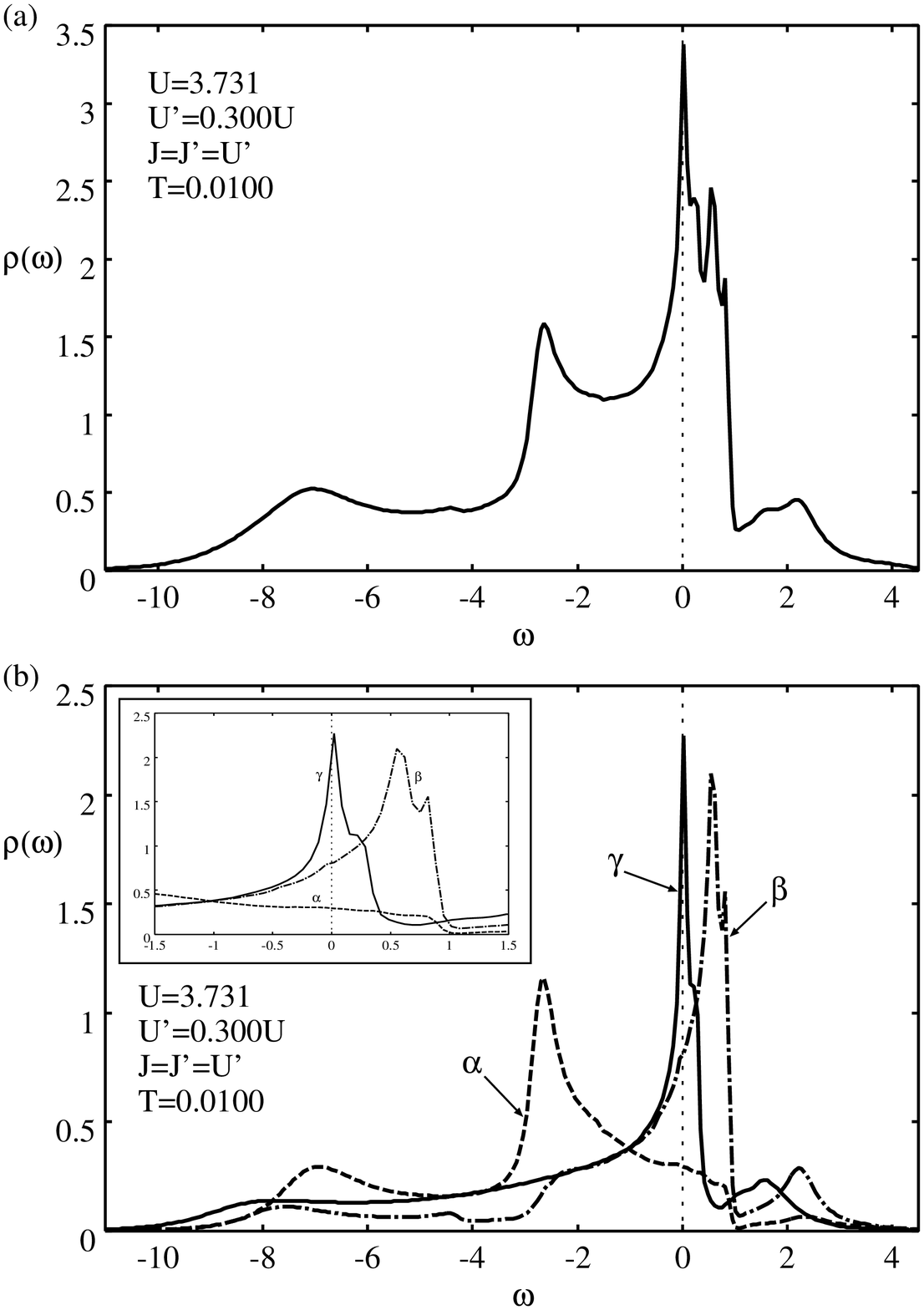}
\caption{T. Nomura \& K. Yamada}
\end{figure}

\clearpage
\begin{figure}
\leavevmode
\epsfxsize=160mm
\epsfbox{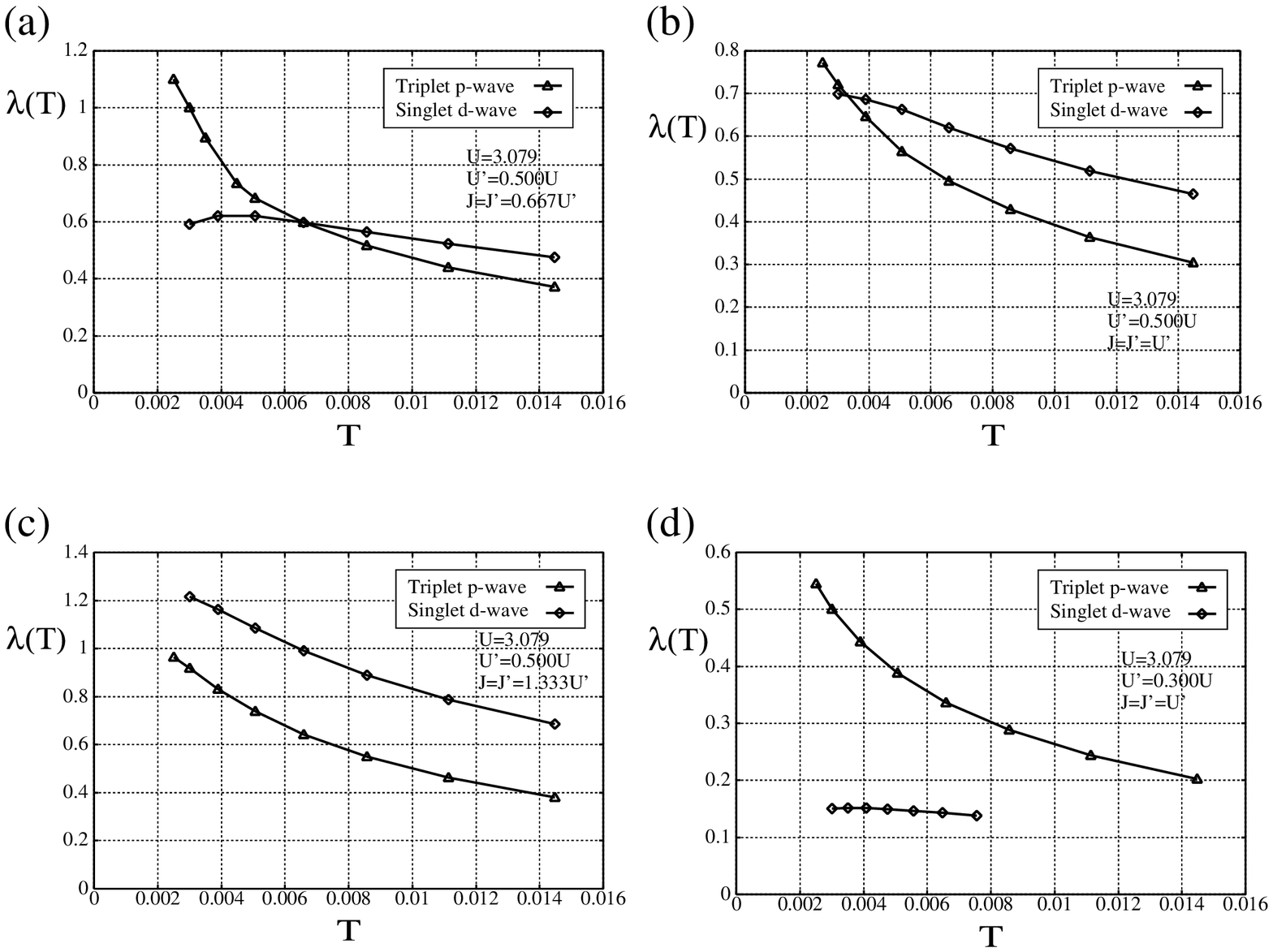}
\caption{T. Nomura \& K. Yamada}
\end{figure}

\clearpage
\begin{figure}
\leavevmode
\epsfxsize=100mm
\epsfbox{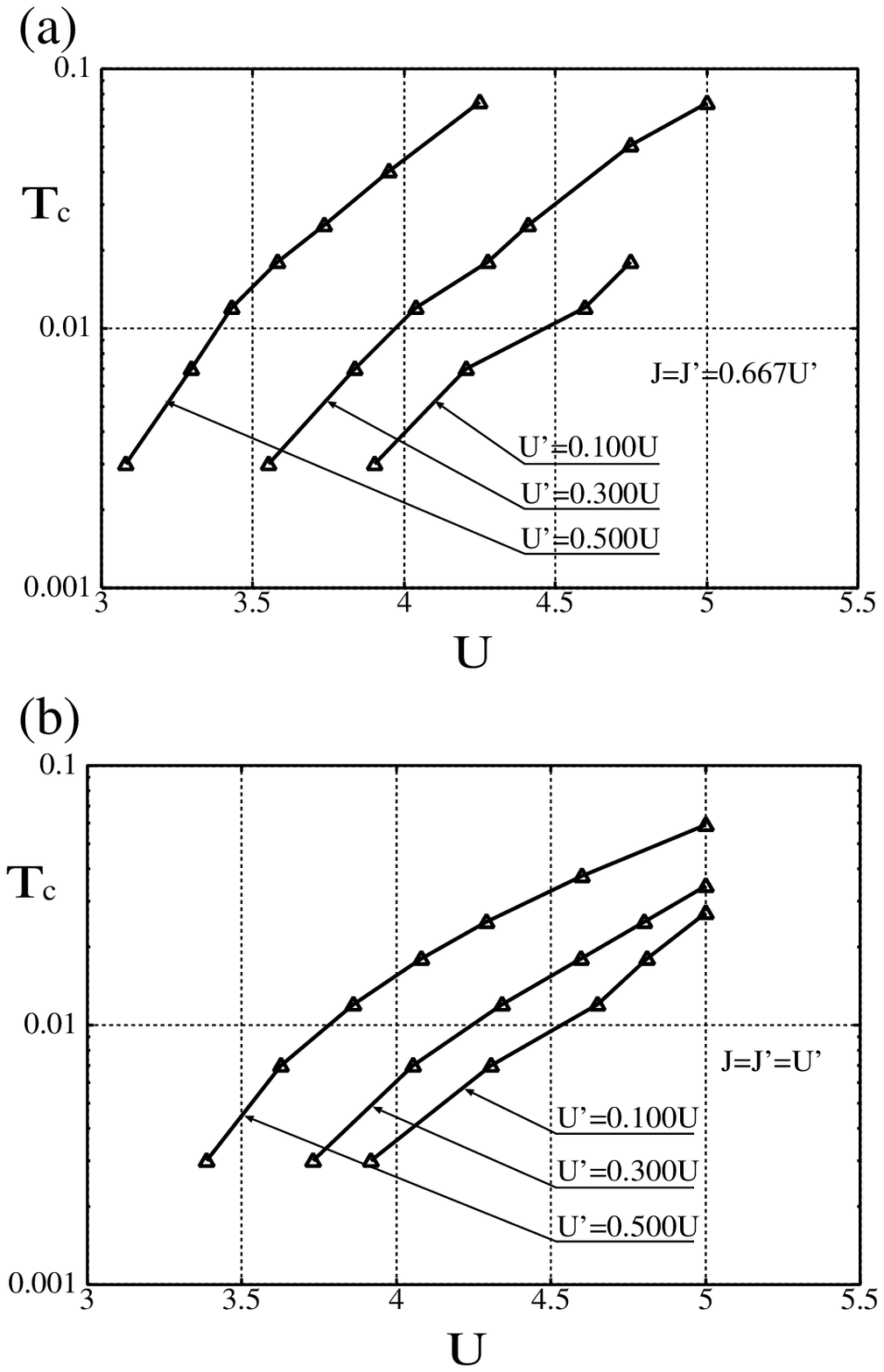}
\caption{T. Nomura \& K. Yamada}
\end{figure}

\clearpage
\begin{figure}
\leavevmode
\epsfxsize=75mm
\epsfbox{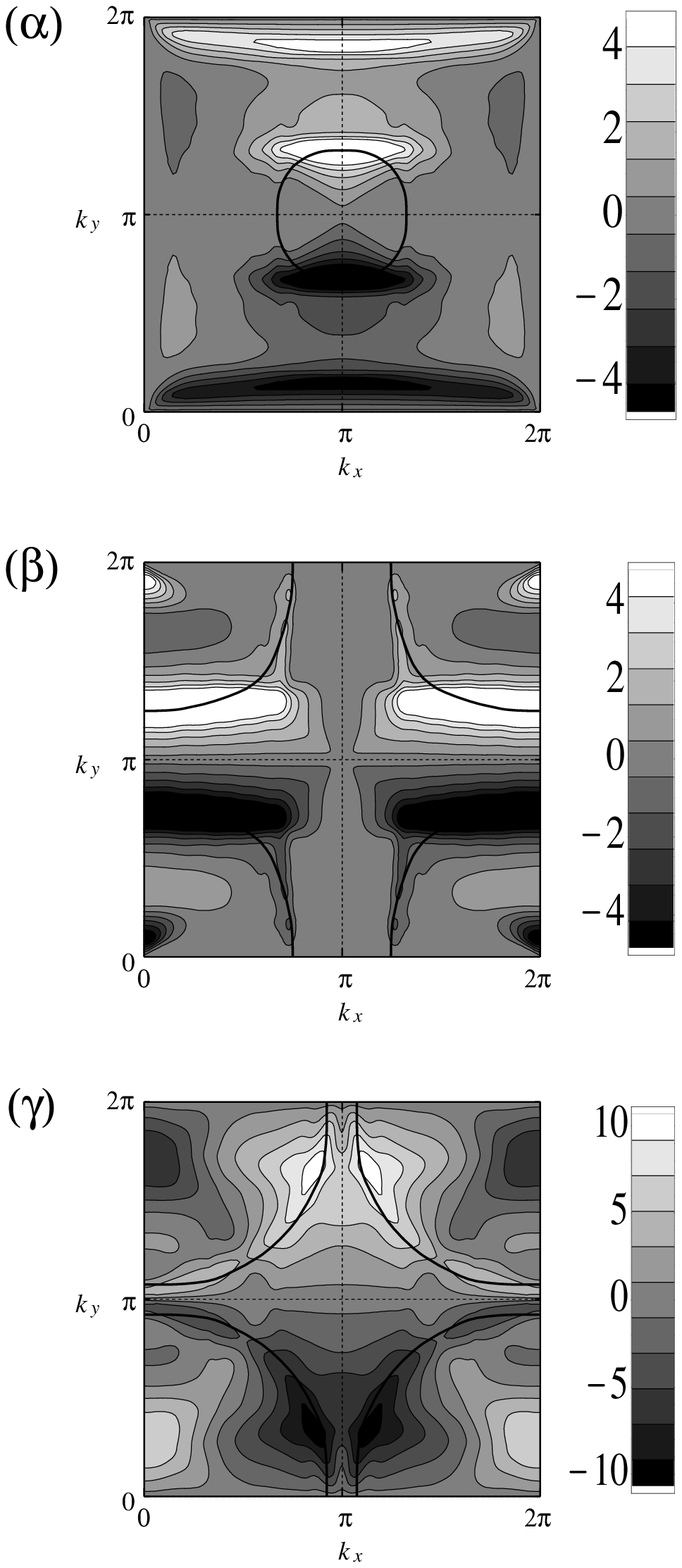}
\caption{T. Nomura \& K. Yamada}
\end{figure}

\clearpage
\begin{figure}
\leavevmode
\epsfxsize=90mm
\epsfbox{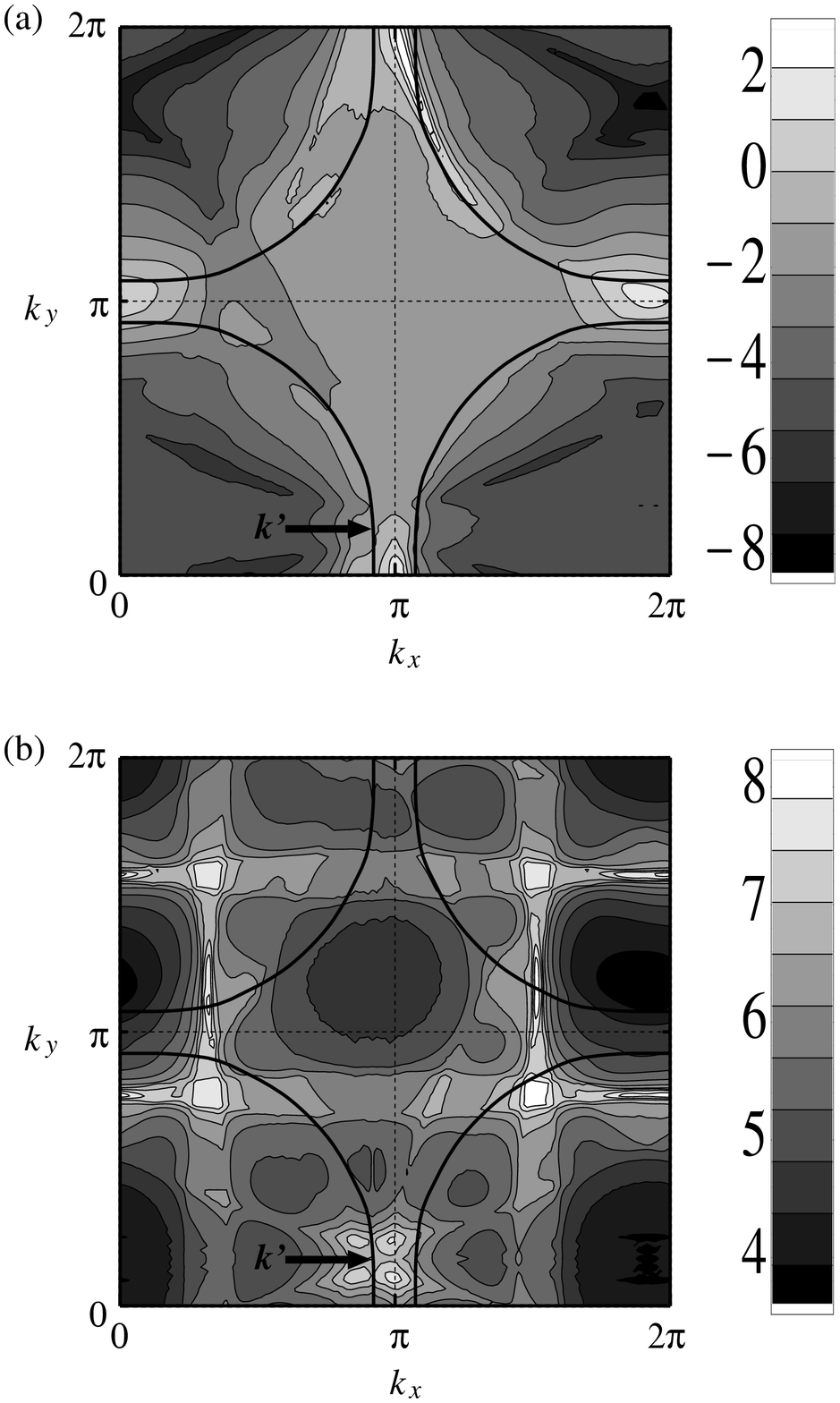}
\caption{T. Nomura \& K. Yamada}
\end{figure}


\begin{thebibliography}{99}
\bibitem{Rf:Maen.1} Y. Maeno, H. Hashimoto, K. Yoshida, S. Nishizaki,
T. Fujita, J. G. Bednorz, and F. Lichtenberg: 
Nature {\bf 372} (1994) 532. 
\bibitem{Rf:Maen.2} Y. Maeno, T. M. Rice and M. Sigrist: 
Phys. Today {\bf 54} (2001) 42. 
\bibitem{Rf:Ishi.1} K. Ishida, H. Mukuda, Y. Kitaoka, K. Asayama, 
Z. Q. Mao ,Y. Mori and Y. Maeno: 
Nature {\bf 396} (1998) 658. 
\bibitem{Rf:Duff.1} J. A. Duffy, S. M. Hayden, Y. Maeno, 
Z. Q. Mao, J. Kulda and G. J. McIntyre: 
Phys. Rev. Lett. {\bf 85}  (2000) 5412. 
\bibitem{Rf:Mack.1} A. P. Mackenzie, R. K. W. Haselwimmer, A. W. Tyler, 
G. G. Lonzarich, Y. Mori, S. Nishizaki and Y. Maeno: 
Phys. Rev. Lett. {\bf 80} (1998) 161. 
\bibitem{Rf:Ishi.2} K. Ishida, H. Mukuda, Y. Kitaoka, 
Z. Q. Mao, Y. Mori and Y. Maeno: 
Phys. Rev. Lett. {\bf 84} (2000) 5387. 
\bibitem{Rf:Luke.1} G. M. Luke, Y. Fudamoto, K. M. Kojima, 
M. I. Larkin, J. Merrin, B. Nachumi, Y. J. Uemura, Y. Maeno, 
Z. Q. Mao, Y. Mori, H. Nakamura and M. Sigrist: 
Nature {\bf 394} (1998) 558. 
\bibitem{Rf:Mao.1} Z. Q. Mao, Y. Maeno, S. NishiZaki, 
T. Akima and T. Ishiguro: 
Phys. Rev. Lett. {\bf 84} (2000) 991. 
\bibitem{Rf:Nish.1} S. NishiZaki, Y. Maeno and Z. Q. Mao: 
J. Phys. Soc. Jpn. {\bf 69} (2000) 572. 
\bibitem{Rf:Maen.3} Y. Maeno, K. Yoshida, H. Hashimoto, 
S. Nishizaki, S. Ikeda, M. Nohara, T. Fujita, A. P. Mackenzie, 
N. E. Hussey, J. G. Bednorz and F. Lichtenberg: 
J. Phys. Soc. Jpn. {\bf 66} (1997) 1405. 
\bibitem{Rf:Mack.2} A. P. Mackenzie, S. R. Julian, A. J. Diver, 
G. J. McMullan, M. P. Ray, G. G. Lonzarich, 
Y. Maeno, S. Nishizaki and T. Fujita: 
Phys. Rev. Lett. {\bf 76} (1996) 3786. 
\bibitem{Rf:Oguc.1} T. Oguchi: 
Phys. Rev. B {\bf 51} (1995) 1385. 
\bibitem{Rf:Sing.1} D. J. Singh: 
Phys. Rev. B {\bf 52} (1995) 1358. 
\bibitem{Rf:Rice.1} T. M. Rice and M. Sigrist: 
J. Phys: Condens. Matter {\bf 7} (1995) L643. 
\bibitem{Rf:Mazi.1} I. I. Mazin and D. J. Singh: 
Phys. Rev. Lett. {\bf 79} (1997) 733. 
\bibitem{Rf:Mont.1} P. Monthoux and G. G. Lonzarich: 
Phys. Rev. B {\bf 59} (1999) 14598. 
\bibitem{Rf:Sidi.1} Y. Sidis, M. Braden, P. Bourges, 
B. Hennion, S. Nishizaki, Y. Maeno and Y. Mori: 
Phys. Rev. Lett. {\bf 83} (1999) 3320. 
\bibitem{Rf:Kuwa.1} T. Kuwabara and M. Ogata:  
Phys. Rev. Lett. {\bf 85} (2000) 4586.
\bibitem{Rf:Kuro.1} K. Kuroki, M. Ogata, R. Arita and H. Aoki: 
Phys. Rev. B {\bf 63} (2001) 060506. 
\bibitem{Rf:Serv.1} F. Servant, B. Fak, S. Raymond, J. P. Brison, 
P. Lejay and J. Flouquet: 
Phys. Rev. B {\bf 65} (2002) 184511. 
\bibitem{Rf:Mina.1} M. Minakata and Y. Maeno: 
Phys. Rev. B {\bf 63} (2001) 180504(R). 
\bibitem{Rf:Brad.1} M. Braden, O. Friedt, Y. Sidis, 
P. Boyrges, M. Minakata and Y. Maeno: 
cond-mat/0107579. 
\bibitem{Rf:Kiku.1} N. Kikugawa and Y. Maeno: Preprint. 
\bibitem{Rf:Taki.1} T. Takimoto: 
Phys. Rev. B {\bf 62} (2000) R14641. 
\bibitem{Rf:Bask.1} G. Baskaran: 
Physica B: {\bf 223 \& 224} (1996) 490. 
\bibitem{Rf:Ng.1} K. K. Ng and M. Sigrist: 
Euro. Phys. Lett. {\bf 49} (2000) 473. 
\bibitem{Rf:Spal.1} J. Spa{\l}ek: 
Phys. Rev. B {\bf 63} (2001) 104513. 
\bibitem{Rf:Nomu.1} T. Nomura and K. Yamada: 
J. Phys. Soc. Jpn. {\bf 69} (2000) 3678. 
\bibitem{Rf:Nomu.2} T. Nomura and K. Yamada: 
J. Phys. Chem. Solids {\bf 63} (2002) 1337. 
\bibitem{Rf:Dagotto.1} For example, E. Dagotto, T. Hotta and A. Moreo: 
Phys. Rep. {\bf 344} (2001) 1. 
\bibitem{Rf:Nomu.3} T. Nomura and K. Yamada: 
J. Phys. Soc. Jpn. {\bf 69} (2000) 1856. 
\bibitem{Rf:Dama.1} 
A. Damascelli, K. M. Shen, D. H. Lu, N. P. Armitage, 
F. Ronning, D. L. Feng, C. Kim, Z.-X. Shen, T. Kimura, 
Y. Tokura, Z. Q. Mao and Y. Maeno: 
J. Electron Spectr. Relat. Phenom. {\bf 114} (2001) 641. 
\bibitem{Rf:Agte.1} D. F. Agterberg, T. M. Rice and M. Sigrist: 
Phys. Rev. Lett. {\bf 78} (1997) 3374. 
\bibitem{Rf:Mazi.2} I. I. Mazin and D. J. Singh: 
Phys. Rev. Lett. {\bf 82} (1999) 4324. 
\bibitem{Rf:Sigr.1} M. Sigrist, D. Agterberg, A. Furusaki, 
C. Honerkamp, K. K. Ng, T. M. Rice and M. E. Zhitomirsky:
Phisica C {\bf 317-318} (1999) 134.
\bibitem{Rf:Bona.1} I. Bonalde, B. D. Yanoff, M. B. Salamon,
D. J. Van Harlingen and E. M. E. Chia: 
Phys. Rev. Lett. {\bf 85} (2000) 4775.
\bibitem{Rf:Lupi.1} C. Lupien, W. A. MacFarlane, C. Proust, 
L. Taillefer, Z. Q. Mao and Y. Maeno:
Phys. Rev. Lett. {\bf 86} (2001) 5986.
\bibitem{Rf:Nomu.4} T. Nomura and K. Yamada: 
J. Phys. Soc. Jpn. {\bf 71} (2002) 404. 
\bibitem{Rf:Okuda.1} N. Okuda, T. Suzuki, Z. Q. Mao, Y. Maeno and T. Fujita: 
J. Phys. Soc. Jpn. {\bf 71} (2002) 1134; 
Physica B, {\bf 312-313} (2002) 800.
\end{thebibliography}
\end{document}